\documentclass[10pt,journal,final,twocolumn]{IEEEtran}

\IEEEoverridecommandlockouts

\usepackage{amsmath,amssymb,amsfonts}
\usepackage{cite}
\usepackage{subfigure}
\usepackage{graphicx}
\usepackage{textcomp}
\usepackage{xcolor}
\usepackage{times}
\usepackage{subfigure}

\usepackage{amssymb,amsmath}
\usepackage{booktabs}

\usepackage{acronym}  

\usepackage{balance}

\usepackage{epstopdf}

\usepackage{bm}   

\usepackage{algorithm} 

\usepackage{algorithmic}

\usepackage{arydshln}

\usepackage{lipsum}
\usepackage{stfloats}
\usepackage{url}

\usepackage{setspace}
\usepackage{hyperref}

\usepackage{xr}
\hypersetup{
	colorlinks = true, %
	linkcolor = black, %
	urlcolor = blue,%
	citecolor = blue} %
\allowdisplaybreaks[4]

\newtheorem{theorem}{\bf Theorem}

\newtheorem{definition}{\bf Definition}
\newtheorem{remark}{\bf Remark}

\newtheorem{lemma}{\bf Lemma}
\newtheorem{corollary}{\bf Corollary}
\newcommand{\mb}[1]{{  \mathbf  #1}}  

\definecolor{BLUE}{rgb}{0,0,1}

\acrodef{siso}[SISO]{single-input single-output}%

\acrodef{nmse}[NMSE]{normalized mean square error}%

\acrodef{ris}[RIS]{reconfigurable intelligent surface}%

\acrodef{mm}[MM]{majorization-minimization}%
\acrodef{IF}[IF]{ice filling}%

\acrodef{rhs}[RHS]{reconfigurable holographic surface}%

\acrodef{csi}[CSI]{channel state information}%
\acrodef{awgn}[AWGN]{additive white Gaussian noise}%

\acrodef{cs}[CS]{compressed sensing}%
\acrodef{ao}[AO]{alternaing optimization}%

\acrodef{bs}[BS]{base station}%
\acrodef{snr}[SNR]{signal-to-noise ratio}%
\acrodef{mmwave}[mmWave]{millimeter-wave}%
\acrodef{snr}[SNR]{signal-to-noise ratio}%
\acrodef{rf}[RF]{radio frequency}%

\acrodef{das}[DAS]{dense array system}%

\acrodef{omp}[OMP]{orthogonal matching pursuit}%
\acrodef{mp}[MP]{message passing}%

\acrodef{ml}[ML]{maximum likelihood}%

\acrodef{mse}[MSE]{mean square error }%

\acrodef{sbr}[S-BAR]{successive Bayesian reconstructor}%

\acrodef{qidd}[QIDD]{quasi-infinite dimension disaster}%

\acrodef{as}[AS]{antenna selection}%

\acrodef{vamp}[VAMP]{vector approximate message passing}%

\acrodef{fas}[FAS]{fluid antenna system}%

\acrodef{swipt}[SWIPT]{simultaneous wireless information and power transfer}%
\acrodef{sinr}[SINR]{signal-to-interference-plus-noise ratio}%

\acrodef{rc}[RC]{reflection coefficient}%
\acrodef{mimo}[MIMO]{multiple-input multiple-output}%
\acrodef{miso}[MISO]{multiple-input single-output}%

\makeatletter
\def\firstletterparse#1#2&{\def\strfirstletter{#1}\def\strotherletters{#2}}

\newcommand{\MakeSmallcaps}[1]{%
	\expandafter\firstletterparse#1&
	\expandafter\MakeUppercase\strfirstletter\textsc{\strotherletters}%
}

\makeatother

\def\BibTeX{{\rm B\kern-.05em{\sc i\kern-.025em b}\kern-.08em
		T\kern-.1667em\lower.7ex\hbox{E}\kern-.125emX}}

\begin{document}
	\title{Ice-Filling: Near-Optimal Channel Estimation for Dense Array 
		Systems
	}
	\author{
		{{Mingyao Cui},~\IEEEmembership{Graduate Student Member,~IEEE},
			{{Zijian Zhang},~\IEEEmembership{Graduate Student Member,~IEEE}}, \\
			{{Linglong Dai},~\IEEEmembership{Fellow,~IEEE},} and
			{{Kaibin Huang},~\IEEEmembership{Fellow,~IEEE}}
		}
		\vspace{-3em}
		\thanks{Mingyao Cui and Kaibin Huang are with  the Department of 
		Electrical and 
			Electronic Engineering, The University of Hong Kong, Hong Kong 
			(e-mails: 
			mycui@eee.hku.hk, huangkb@eee.hku.hk).}
		\thanks{Zijian Zhang and Linglong Dai are with the Department of 
		Electronic 
			Engineering, Tsinghua University, and the State Key laboratory of 
			Space 
			Network 
			and Communications, Tsinghua University, Beijing 100084, China. 
			(e-mails: 
			zhangzj20@mails.tsinghua.edu.cn; daill@tsinghua.edu.cn.)}
	}	
	\maketitle
	\begin{abstract}
		By deploying a large number of antennas with sub-half-wavelength 
		spacing in 
		a compact space, dense array systems (DASs) can fully unleash the 
		multiplexing and diversity gains of limited apertures. To acquire these 
		gains, accurate channel state information acquisition is 
		necessary but challenging due to the large antenna numbers. 
		To 
			overcome 
			this obstacle, this paper reveals that designing the observation 
			matrix to exploit the high spatial correlation of DAS channels is 
			crucial 
			for realizing near-optimal Bayesian channel estimation.	
		Specifically, we 
		prove that the observation matrix design for channel 
		estimation 
		is equivalent to a time-domain duality of point-to-point multiple-input 
		multiple-output precoding, except for the change in the total power 
		constraint on the precoding matrix to the pilot-wise discrete power 
		constraint on the observation matrix. Inspired by 
		Bayesian 
		regression,
			a novel ice-filling algorithm is proposed to design 
			amplitude-and-phase 
			controllable observation matrices, 
		and a majorization-minimization algorithm is proposed to address the 
		phase-only controllable case. Particularly, we prove that the 
		ice-filling 
		algorithm can be interpreted as a ``quantized'' water-filling 
		algorithm, 
		wherein 
		the latter's continuous power-allocation process is converted into the 
		former's discrete pilot-assignment process.   
		To support the near-optimality of the proposed designs, we provide 
		comprehensive analyses on the achievable mean square errors and 
		their asymptotic expressions. Finally, numerical results confirm that 
		our proposed designs achieve the near-optimal channel estimation 
		performance and outperform existing approaches significantly.
		
	\end{abstract}
	
	\section{Introduction}
	
	From 3G to 5G, antenna array systems play an irreplaceable role in wireless 
	communications \cite{ZhangJiaYi'20,Shlezinger'21}.  
	By strategically configuring multiple antennas, an antenna array can 
	constructively manipulate the radiated signals, allowing for enhanced 
	wireless 
	coverage, improved spectral efficiency, and reduced power consumption 
	\cite{HanChong'21,Heath'23,Gaozhen'23}. 
	The performance of arrays continuously improves with the number of antennas 
	\cite{Cui'22'TCOM,CuimingYao'23'Mag}. However, in practical scenarios, the 
	space available for antenna deployment is usually limited, which restricts 
	the 
	performance gains endowed by antenna arrays \cite{Zijian'23'JSAC}. To 
	achieve a
	better performance with a spatially-limited aperture, 
	\acp{das} have attracted extensive attentions in recent years 
	\cite{liu2023densifying}. 
	
	Generally, \acp{das} represent a series of array technologies that massive 
	sub-wavelength antennas are densely arranged within a compact space. Unlike 
	the 
	conventional arrays with half-wavelength antenna spacing $\lambda/2$, the 
	antenna spacing of \acp{das} is much smaller, such as $\lambda/6$ 
	\cite{Ovejero'17'TAP}, $\lambda/8$ \cite{hwang2020binary}, $\lambda/10$ 
	\cite{Christos'18'COMMAG}, or even $\lambda/23$ \cite{liu2019deeply}. With 
	an 
	increased number of sub-channels, \acp{das} promise to achieve high array 
	gains 
	and 
	fully exploit the multiplexing and diversity gains of limited apertures 
	\cite{liu2023densifying}. Besides, \acp{das} can also reduce the grating 
	lobes 
	and provide high performance for large values of oblique angles of 
	incidence 
	\cite{di2023mimo}. The 
	typical \ac{das} realizations include holographic \ac{mimo} 
	\cite{HuangChongWen'20'WCOM}, \acp{ris} 
	\cite{zhang2023reconfigurable}, fluid antenna systems (FASs) 
	\cite{wong2020fluid,zhang2023successive}, graphene-based 
	nano-antenna arrays \cite{Hanchong'18'VTC, Hanchong2017Graphene},
	radio-frequency lens \cite{CBC2016RFlen}, and so on. 
	For example, in 
	\cite{Ovejero'17'TAP}, massive sub-wavelength patches of 
	$\lambda/6$-spacing 
	are 
	densely printed on a holographic \ac{mimo} surface to generate multiple 
	beams 
	flexibly. In \cite{yurduseven2018dynamically}, meta-elements sized of 
	$\lambda/2.5\times\lambda/17.5$ are seamlessly integrated onto a feeding 
	microstrip to achieve \ac{rhs} beam steering. In \cite{liu2019deeply}, a 
	\ac{ris} composed of antennas sized of $\lambda/23\times\lambda/23$ is 
	designed 
	and fabricated for terahertz communications. 
	
	The performance gains of antenna arrays are realized via the constructive 
	beamformers enabled by their phase-shifters (PSs) and \ac{rf} chains. To 
	realize 
	effective beamforming, the 
	acquisition of accurate \ac{csi} is essential, particularly in situations 
	where 
	the number of RF chains is smaller than the number of massive antennas. 
	{\color{black} Up to 
		now, many 
		estimators have 
		been 
		proposed to acquire the \ac{csi} of large antenna arrays, and most of 
		them can 
		be adopted in 
		DASs~\cite{Gaozhen'18'WCOM,Ziwei'20'TVT,lee2016channel,Malong'20'TCSP,Chongwen'19'TSP,
			rangan2019vector,Zhu2021NearCE,gonzalez2020wideband,ma2020data,CNNAMP_Hu2023,Kwan2024DLCE,
			Alkhateeb2015BT, Xiao2016BT,Tianyue2024CBT,Xisuo'21'JSAC}. Their 
			fundamental 
		principle 
		involves receiving pilot signals using observation matrices  
		and recovering the channel by advanced estimation algorithms.}
	For example, when the available pilot length is larger than the antenna 
	number, the classical least square (LS) estimator is usually adopted. 
	Leveraging 
	the property of 
	channel sparsity, compressed sensing (CS)-based channel estimators are 
	widely 
	studied to improve the estimation accuracy and reduce the pilot overhead
	\cite{Gaozhen'18'WCOM,Ziwei'20'TVT}, such as the \ac{omp}-based estimator 
	\cite{lee2016channel,Malong'20'TCSP}, the \ac{mp}-based estimator 
	\cite{Chongwen'19'TSP,rangan2019vector,Zhu2021NearCE}, 
	and the gridless 
	sparse 
	signal reconstructor \cite{gonzalez2020wideband}. By 
	training neural networks with a large amounts of channel data, the deep 
	learning approaches are also utilized to realize both data-driven and 
	model-driven channel estimators 
	\cite{ma2020data,CNNAMP_Hu2023,Kwan2024DLCE, Xisuo'21'JSAC}. Besides,  beam 
	alignment 
	techniques 
	\cite{Alkhateeb2015BT, Xiao2016BT,Tianyue2024CBT}, including  
	beam 
	sweeping and hierarchical beam training,  have also been widely 
	explored to acquire the implicit CSI with low pilot overhead. 
	
	
	Although most of existing channel estimators can be adopted in 
	\acp{das}, they fail to fully exploit the high spatial correlation 
	of \ac{das} channels, thereby leaving a remarkable performance gap from the 
	optimal estimator. 
	Specifically, this spatial correlation is attributed to the fact that, 
	the extremely-dense deployment of \ac{das} antennas significantly increases 
	the similarity of radio waves impinging on antenna ports and aggravates the 
	mutual-coupling effect between adjacent antenna circuits. 
	This fact makes the covariance 
	matrices 
	of \ac{das} channels no longer diagonal but highly 
	structured. 
	For diagonal covariance matrices, the observation matrices for receiving 
	pilot signals are either generated randomly or set as predefined codebooks, 
	such as the Discrete Fourier Transform (DFT) 
	matrix~\cite{Gaozhen'18'WCOM,Ziwei'20'TVT,lee2016channel,gonzalez2020wideband,CNNAMP_Hu2023,
		Chongwen'19'TSP,Malong'20'TCSP,rangan2019vector,ma2020data,Xisuo'21'JSAC,Alkhateeb2015BT,
		Xiao2016BT,Zhu2021NearCE,Kwan2024DLCE, Tianyue2024CBT}, which are 
	sufficient to achieve the optimality of channel estimation 
	\cite{tsaig2006extensions,donoho2006compressed}.  
	In contrast, since the correlation matrix of \ac{das} channels are full of 
	structural 
	features, it is believed that their observation matrices can be tightly 
	aligned with these features in pilot transmission. 
	This structured 
	pilot-sensing process has a high potential to remarkably boost the channel 
	estimation accuracy in DASs 
	\cite{williams1995gaussian,Srinivas2012GPR,schulz2018tutorial}. 
	{\color{black} 
		For example, in \cite{zhang2023successive}, the structured channel 
		covariance 
		is utilized to determine the positions of fluid antennas during the 
		pilot 
		reception. Due to the strong channel correlation of FAS channels, the 
		estimator 
		in \cite{zhang2023successive} can achieve higher accuracy than CS 
		methods. 
		Authors in \cite{Xubowen_2024} further 
		extend the method in \cite{zhang2023successive} to sparse channel 
		scenarios. 
		In 
		\cite{Emil'22}, aiming at efficient CSI acquisition in holographic MIMO 
		systems, the electromagnetic channel covariance is modeled and utilized 
		to 
		enable a Bayesian channel estimator. However, 
		these existing 
		estimators often rely on specific array structures or parametric 
		channel 
		models, which can neither be used to design the general observation 
		matrix in 
		DASs 
		nor guarantee the performance to be optimal.
	}
	
	
	To fill in this blank, our work represents the first attempt on designing 
	and 
	analyzing the near-optimal observation matrices in the \ac{das} channel 
	estimation. {\color{black} The classical Bayesian regression method is 
	adopted as 
		the channel 
		estimator, i.e., the minimum-mean-square-error (MMSE) method, while our 
		key 
		contributions lie in the designed observation matrices 
		to enhance Bayesian regression schemes, as summarized below.  }
	\begin{itemize}
		\item \textbf{Mutual-information-maximization (MIM) based framework}: 
		From 
		the perspective of MIM, we propose to design the \ac{das}'s observation 
		matrix by minimizing the information 
		uncertainty between the received pilots and wireless channels.
		The 
		formulated MIM problem for observation matrix design is shown to be a 
		time-domain 
		duality of point-to-point MIMO precoding design. In particular, the 
		multi-RF-chain 
		resources for MIMO precoding is transformed to the multi-timeslot-pilot 
		resources for channel estimation, and the total power 
		constraint on the precoding matrix is switched to the pilot-wise 
		discrete 
		power constraint on the observation matrix. Motivated by this finding, 
		we 
		show that the ideal (but unachievable in practice) observation matrix 
		could be constructed by the water-filling principle, which involves the 
		use 
		of the 
		eigenvectors 
		of channel covariance matrix as well as optimal power 
		allocation. 
		
		\item \textbf{Ice-filling enabled observation matrix design}: 
		For practical DAS channel estimations, the continuous power-allocation 
		by 
		water-filling is not 
		applicable as the pilot length is discrete and each pilot transmission 
		has independent power constraint. To deal with this issue, we propose 
		an 
		ice-filling algorithm to design the amplitude-and-phase 
		controllable observation matrices. {\color{black} Inspired by Bayesian 
		regression,
			this algorithm sequentially generates the optimal blocks of the 
			observation 
			matrix by maximizing the mutual information (MI) increment between 
			two 
			adjacent pilot transmissions.} 
		An important insight is that the ice-filling algorithm converts 
		the 
		\emph{power-allocation-process} of the ideal water-filling algorithm 
		into 
		an 
		\emph{eigenvector-assignment-process}. We rigorously prove 
		that, the continuous powers 
		allocated by water-filling 
		are quantized as the number of times that {\it each eigenvector of the 
		channel covariance is assigned for pilot transmission} by ice-filling, 
		with a quantization 
		error smaller than one. The proven 
		quantization 
		nature thereby ensures the near-optimality of the proposed ice-filling 
		algorithm.

		\item \textbf{Majorization minimization (MM) enabled observation matrix 
			design}: We then apply our framework for DAS channel estimation to 
			the 
		situation when only the phases of 
		the 
		observation matrix's weights are controllable while their amplitudes 
		are  
		fixed, making both the water-filling and ice-filling
		algorithms 
		invalid. 
		To address this challenge, we propose a MM algorithm for designing the 
		observation 
		matrix. Its novelty lies in replacing the 
		primal non-convex MIM problem with a series of tractable approximate 
		subproblems having analytical solutions, 
		which are solved in an alternating optimization manner. 
		Numerical simulations demonstrate that 
		its channel estimation performance is comparable to the ice-filling 
		algorithm.
		

		%
		%
		
		\item \textbf{ Performance analysis}:
		Comprehensive analyses on the achievable mean square errors (MSEs) are 
		provided 
		to validate the 
		effectiveness 
		of the proposed designs. We prove that, compared to the 
		random observation matrix design, the ice-filling algorithm 
		can significantly improve the estimation accuracy by the order of the 
		ratio between the number of antennas and the rank of channel 
		covariance. 
		Furthermore, based on the derived quantization nature of ice-filling, 
		the MSE gap between water-filling and ice-filling is proved to decay 
		quadratically with the pilot length, demonstrating the near-optimality 
		of the 
		ice-filling algorithm. In addition, the close-form MSEs under imperfect 
		channel 
		covariance are also derived. The superiority of the proposed algorithms 
		is 
		further validated using extensive numerical results. 
		
		
		

	\end{itemize}
	
	The remainder of the paper is organized as follows. The system model is 
	presented in Section \ref{sec:2}. The proposed MIM-based framework for 
	observation matrix design is given in Section \ref{sec:3}. The 
	ice-filling and MM algorithms are elaborated in Section \ref{sec:4}. 
	Then, 
	the MSEs are analyzed in Section \ref{sec:5}. 
	Simulation results are carried out in Section \ref{sec:6}, and finally the 
	conclusions are drawn in Section \ref{sec:7}.
	
	\textit{Notation:} Lower-case and upper-case boldface letters represent
	vectors and matrices, respectively. ${[\cdot]^{-1}}$, ${[\cdot]^{\dag}}$, 
	${[\cdot]^{*}}$, ${[\cdot]^{\rm 
			T}}$, and ${[\cdot]^{\rm H}}$ denote the inverse, pseudo-inverse, 
			conjugate, 
	transpose, 
	and conjugate-transpose operations, respectively; $\|\cdot\|_2$ denotes the 
	$l_2$-norm of the argument; $\|\cdot\|_F$ denotes the Frobenius norm of the 
	argument; 
	$|\cdot|$ denotes the element-wise amplitude of its argument; 
	$\mathsf{Tr}(\cdot)$ denotes the trace operator; 
	${\mathsf E}\left(\cdot\right)$ is the expectation operator; 
	$\Re\{\cdot\}$ denotes the real part of the argument; 
	$\lambda_{\max}(\cdot)$
	denotes the largest eigenvalue of its argument; 
	$\mathcal{C} 
	\mathcal{N}\!\left({\bm \mu}, {\bf \Sigma } \right)$ denotes the 
	complex Gaussian distribution with mean ${\bm 
		\mu}$ and covariance ${\bf \Sigma }$; ${\cal U}\left(a,b\right)$ 
		denotes 
	the 
	uniform distribution between $a$ and $b$; $\mathbf{I}_{L}$ is an $L\times 
	L$ 
	identity matrix; $\mathbf{1}_{L}$ is an $L$-dimensional all-one vector; and 
	$\mathbf{0}_{L}$ is an all-zero vector or matrix with dimension $L$.


	\section{System Model}\label{sec:2}

	\begin{figure}[!t]
		\centering
		\includegraphics[width=0.47\textwidth]{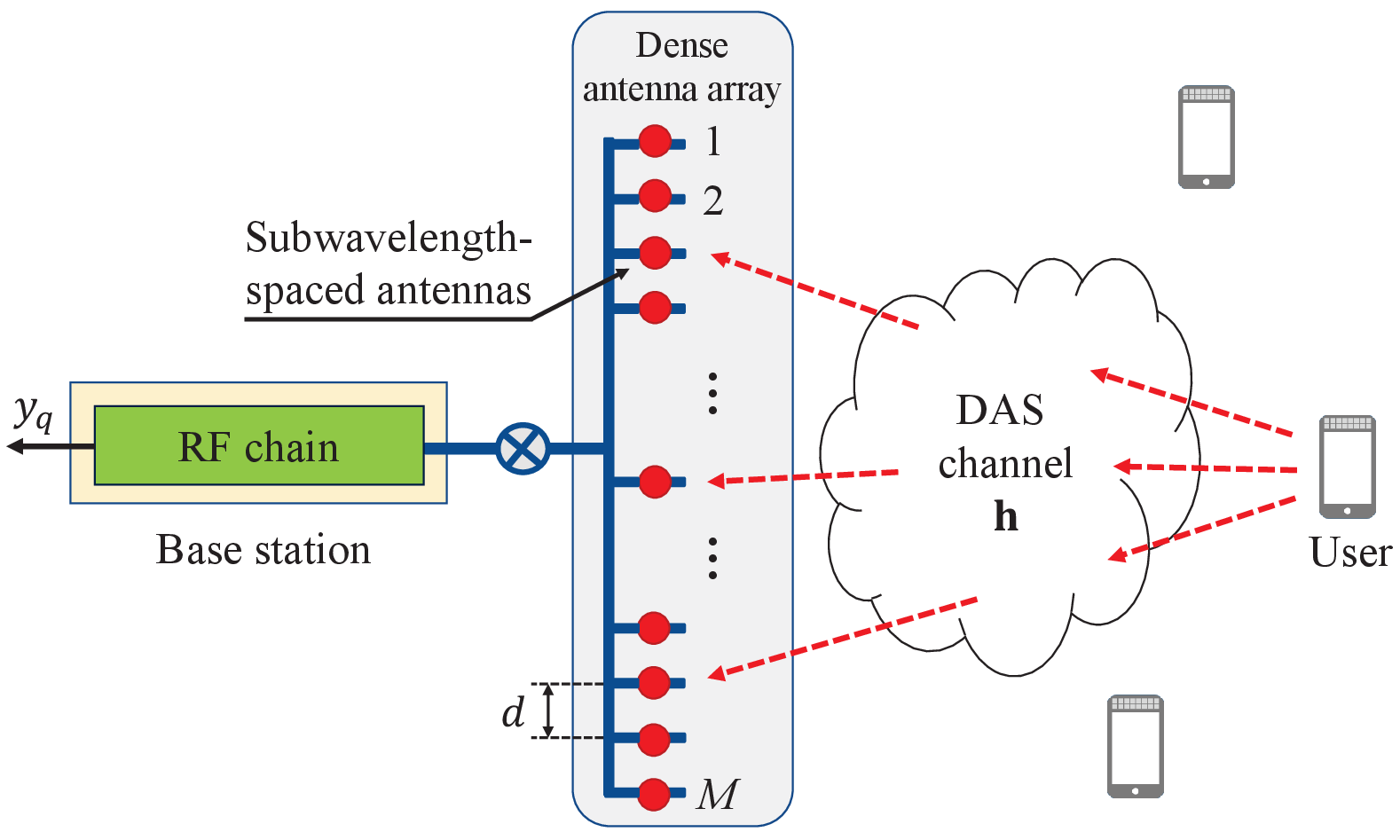}
		\vspace*{-0.5em}
		\caption{An illustration of uplink channel estimation for 
		{\color{black} 
		a} \ac{das}.}
		\vspace*{-1em}
		\label{img:scenario}
	\end{figure}

	As illustrated in Fig. 
	\ref{img:scenario}, we consider the narrowband channel estimation of an 
	uplink 
	single-input 
	multiple-output (SIMO) system. The \ac{bs} employs an $M$-antenna DAS, 
	which 
	comprises an analog combining structure supported by one 
	\ac{rf} chain, to receive 
	the pilots from a 
	single-antenna user. 
	
	Let ${\bf h}\in{\mathbb C}^{M\times 1}$ 
	be the channel vector and $Q$ the number of transmit pilots within a 
	coherence-time frame. The  
	received signal ${y}_q\in{\mathbb C}$  at the \ac{bs} in timeslot $q$ 
	is modeled as 
	\begin{equation}\label{eqn:y_p}
		{y}_q = {\bf w}^H_q{\bf h}s_q + {\bf w}^H_q{\bf z}_q,
	\end{equation}
	where ${\bf w}_q\in{\mathbb C}^{M\times 1}$ is the observation vector at 
	the 
	\ac{bs}, $s_q$ 
	the pilot transmitted by the user, and ${\bf z}_q \sim \mathcal{C} 
	\mathcal{N}\!\left({\bf 0}_{M},  \sigma^2{\bf I}_{M} \right)$ the 
	additive white Gaussian noise. As the observation vector ${\bf w}_q$  
	affects both the desired signal ${\bf h}s_q$ and the noise ${\bf z}_q$, its 
	power has no impact on the estimation accuracy. Therefore, without any loss 
	of 
	generality, we can normalize ${\bf w}_q$ to $\|{\bf w}_q\|^2_2 = 1$, which 
	reshapes the noise distribution to 
	${\bf w}_q^H{\bf z}_q \sim 
	\mathcal{C} 
	\mathcal{N}\!\left(0,  \sigma^2 \right)$. Two practical implementations of 
	the 
	observation vector ${\bf w}_q$ will be discussed in our research: the 
	amplitude-and-phase controllable combiner and the phase-only 
	controllable combiner. The former connects each antenna to the RF chain 
	using 
	one 
	PS and one low noise amplifier (LNA), enabling adjustment of both the  
	amplitude and phase of the coefficients of ${\bf w}_q$. The latter deploys 
	a 
	single PS 
	between each antenna and 
	the RF chain, allowing only phase adjustment of the coefficients of ${\bf 
	w}_q$ 
	and enforcing a 
	constant modulus constraint on the observation vector, i.e., $|w_{q, 
		m}| = \frac{1}{\sqrt{M}}$ for $m \in \{1,2,\cdots, 
	M\}$~\cite{Gaozhen'18'WCOM}.

	Considering the total $Q$-timeslot pilot 
	transmission, the received signal vector ${\bf y} := \left[{ 
		y}_1,\cdots,{y}_Q\right]^{\rm T}$ is expressed as 
	\begin{equation}\label{eqn:y}
		{\bf y} = {\bf W}^{H}{\bf h} + {\bf z}, 
	\end{equation}
	where $s_q$ is assumed 
	to 
	be 1 for all $q\in\{1,\cdots,Q\}$, ${\bf{W}} = 
	\left[{\bf{w}}_1,\cdots, {\bf{w}}_Q\right]$ stands for the observation 
	matrix, 
	and 
	${\bf z} := \left[ {\bf 
		w}^H_1{\bf z}_1,\cdots, {\bf w}^H_Q{\bf z}_Q \right]^{\rm T}$. 
	As the power 
	of $\mb{w}_q$ is normalized to 1, we have ${\bf z}\sim 
	\mathcal{CN}(\mb{0}_Q, \sigma^2 \mb{I}_Q)$.
	The objective of channel estimation is to recover the channel $\bf h$ from 
	the 
	received pilot $\bf y$. As a DAS possesses 
	highly correlated channels across different antenna ports, we adopt 
	Bayesian 
	regression to recover $\mb{h}$. Suppose the channel
	is generated from a Gaussian process $\mathcal{CN}(\mb{0}_M,  
	\bm{\Sigma}_{\bf h})$, where the covariance matrix $\bm{\Sigma}_{\bf h} \in 
	\mathbb{C}^{M\times M}$, also called \emph{prior 
		kernel},  characterizes our prior knowledge to the channel
	correlation. {\color{black}
		The kernel reflects the long-term statistical information of the 
		covariance 
		of channels, i.e., ${\bf \Sigma}={\mathsf E}({\bf h}{\bf h}^H)$, which 
		can 
		keep unchanged for a long time. To obtain 
		the prior kernel in practical MIMO systems, an 
		ideal approach is to leverage the channel realizations generated by 
		some 
		standard models or real-world 
		datasets. In addition, some online second-moment learning methods 
		proposed 
		in 
		\cite{Khalilsarai;20,Sungwoo'18} can also be used to obtain the prior 
		kernel efficiently. 
	} Given the prior kernel, the joint probability distribution of 
	$[\mb{h}^T,\mb{y}^T]^T$ satisfies
	\begin{equation}
		\begin{aligned}
			\mathcal{CN} \left(
			\left[ \begin{array}{c}
				\mb{0}_M \\
				\mb{0}_Q
			\end{array} \right], 
			\left[ 
			{\begin{array}{*{20}{c}}
					{{{\bf{\Sigma }}_{\bf{h}}}}&{{\bf{\Sigma 
						}}_{\bf{h}}}{{\bf{W}}}\\{{\bf{W}}^H}
					{{\bf{\Sigma 
						}}_{\bf{h}}}& {\bf W}^{H}{\bf{\Sigma 
					}}_{\bf{h}}{\bf W} + \sigma^2 \mb{I}_Q
			\end{array}} \right]\right).
		\end{aligned}
	\end{equation}
	{\color{black} 
		By invoking the property of the linear transform of Gaussian random 
		vectors, the  
		posterior mean and posterior covariance of $\mb{h}$ given 
		observation 
		$\mb{y}$ can be calculated as:
		\begin{align}
			{{\bm{\mu }}}_{\mb{h}|\mb{y}} &=  {{\bf{\Sigma 
				}}_{\bf{h}}}{{\bf{W}}}
			{\left( {\bf W}^{H}{\bf{\Sigma 
				}}_{\bf{h}}{\bf W} + \sigma^2 {\bf I}_Q \right)^{ - 
					1}} 
			{{\bf{y}}}, \label{eqn:bayesian}\\
			{{\bf{\Sigma }}}_{\mb{h}|\mb{y}} &= {{\bf{\Sigma 
				}}_{\bf{h}}} -
			{{\bf{\Sigma 
				}}_{\bf{h}}}{{\bf{W}}}
			{\left( {\bf W}^{H}{\bf{\Sigma 
				}}_{\bf{h}}{\bf W} + \sigma^2 {\bf I}_Q \right)^{ - 
					1}}{ 
				{{{\bf{W}}^H}} }
			{{\bf{\Sigma }}_{\bf{h}}}.\label{eq:posterir_cov}
		\end{align}
		The optimal Bayesian estimation to the channel vector is exactly the 
		posterior 
		mean, i.e., $\hat{\mb{h}} = {{\bm{\mu }}}_{\mb{h}|\mb{y}}$, which is 
		also known 
		as the MMSE estimator. 
	}
	Notably, the posterior covariance, ${\bf \Sigma_{\mb{h}|\mb{y}}}$, also 
	called 
	\emph{posterior kernel}, characterizes the channel estimation 
	error and is a function of the prior kernel ${{\bf{\Sigma }}_{\bf{h}}}$ and 
	the 
	observation matrix $\mb{W}$. Due to 
	the extremely high spatial correlation 
	exhibited by DAS channels, the prior kernel ${{\bf{\Sigma }}_{\bf{h}}}$ 
	would 
	remarkably deviate from the identity matrix. This deviation indicates that 
	aligning the observation matrix $\mb{W}$ with the kernel's 
	subspace could be greatly helpful in reducing the estimation error. 
	Motivated by this fact, this paper concentrates on the design of 
	observation 
	matrix $\mb{W}$, by considering both the amplitude-and-phase controllable 
	and 
	phase-only controllable analog combiners, to boost the channel estimation 
	performance. 
	


	\section{Mutual Information Maximization for Channel Estimation in 
	\acp{das}}\label{sec:3}
	In this section, we first {\color{black} present} a MIM based 
	framework for designing observation matrices. Then, a 
	water-filling-inspired 
	observation-matrix-design is investigated as an ideal case of the proposed 
	framework. 
	
	\subsection{MIM based framework for observation matrix design}
	We adopt the MI between the received 
	signal $\mb{y}$ and 
	the channel $\mb{h}$ as the metric to evaluate the quality of the designed 
	observation matrix. 
	There are two reasons for using the MI as a metric. Firstly, MI tells the 
	amount of information about the unknown channel  
	$\mb{h}$ we can gain from the received signal $\mb{y}$.  The second reason 
	is that, according to the information-theoretic properties of Bayesian 
	statistics from \cite{IT_Clarke1990}, the maximization of MI is 
	asymptotically 
	the 
	minimization of Cramer Rao Bound (CRB). 
	
	Given the distributions
	$\mb{h}\sim\mathcal{CN}(\mb{0}_M,  
	\bm{\Sigma}_{\bf h})$ and ${\bf z} \sim\mathcal{CN}(\mb{0}_Q,  
	\sigma^2\mb{I}_Q)$, the MI is formulated as
	\begin{align}\label{eq:MI}
		\max_{\mb{W}\in\mathcal{W}} I(\mb{y}; \mb{h}) 
		&=\log\det\left( \mb{I}_Q +  \frac{1}{\sigma^2}\mb{W}^H{\bf 
			\Sigma_h} \mb{W} \right),
	\end{align} 
	where $I(\cdot;\cdot)$ denotes the MI, $\det(\cdot)$ is
	the determinant of its argument, and $\mathcal{W}$ represents the feasible 
	set 
	of $\mb{W}$ for both amplitude-and-phase controllable and phase-only 
	controllable analog combiners.
	It is notable that the observation matrix 
	design problem 
	in \eqref{eq:MI} resembles the well-know point-to-point MIMO precoding 
	problem. 
	Comprehensive comparisons 
	between 
	these two problems are provided below.

	\begin{itemize}
		\item \textbf{Category of unknown data}: 
		Both MIMO precoding and observation matrix design in \eqref{eq:MI} aim 
		to 
		reduce the 
		uncertainty between the received data, $\mb{y}$, and the unknown data. 
		The former regards the transmitted symbol vector, denoted by 
		$\mb{x}$, as the unknown data. To facilitate MIMO precoding, 
		\emph{multiple 
			\ac{rf} chains} are employed to 
		support an  equal number of data streams. 	
		On the other hand, the unknown data to the latter is the  
		channel, $\mb{h}$. Our considered uplink channel estimation requires 
		\emph{transmitting pilot signals over multiple timeslots} to recover 
		$\mb{h}$. In comparison, one can discover that the multiple RF chains 
		in MIMO precoding are converted to the multi-timeslot pilot resources 
		in  
		channel estimation. Thereby, the design of the observation 
		matrix in \eqref{eq:MI} can be viewed as a 
		time-domain 
		duality of point-to-point MIMO precoding. 
		
		\item \textbf{Constraint on the observation matrix}: 
		The 
		mathematical difference between MIMO precoding and observation matrix 
		design for uplink channel estimation is attributed to the constraints 
		of $\mb{W}$. For MIMO 
		precoding, each column of $\mb{W}$ refers to a precoding vector 
		associated with one dedicated RF chain. The optimization of precoding 
		vectors, 
		leveraging multiple RF chains, 
		typically enforces  
		the \emph{total-power-constraint} on 
		$\mb{W}$, i.e.,  $\|\mb{W}\|_F^2 = Q$. In terms of 
		the observation matrix design, each column of $\mb{W}$ signifies an 
		observation  
		vector used for receiving one pilot signal. 
		As discussed in Section~\ref{sec:2}, since each observation vector 
		$\mb{w}_q$ amplifies 
		both the desired signal and the noise, it is subject 
		to the \emph{pilot-wise power constraint}, i.e., 
		$\|\mb{w}_q\|_2^2 = 1$ or 
		$|w_{q,m}| = 
		\frac{1}{\sqrt{M}}$. The distinction in 
		the 
		observation matrix constraint differentiates  the MIM in \eqref{eq:MI} 
		from the 
		classical MIMO precoding. 
	\end{itemize}
	

	\subsection{Water-filling Inspired Ideal Observation Matrix Design}
	Prior to addressing problem (\ref{eq:MI}), we would like to consider an 
	ideal (but practically unachievable) situation to establish an upper bound 
	of 
	(\ref{eq:MI}). Specifically, we temporarily make the ideal 
	assumption that the noise vector $\mb{z}$ in \eqref{eqn:y} is independent 
	of 
	the observation matrix $\mb{W}$ and its distribution is always 
	$\mathcal{CN}(\mb{0}_Q, 
	\sigma^2\mb{I}_Q)$ regardless of 
	the power of $\mb{w}_q$. In this context, the pilot-wise power 
	constraint 
	$\|\mb{w}_q\|^2_2 = 1$ can be relaxed to the 
	total power constraint $\|\mb{W}\|_F^2 = \sum_{q = 1}^Q 
	\|\mb{w}_q\|^2_2 = Q$, and problem 
	(\ref{eq:MI}) becomes identical to the point-to-point MIMO precoding 
	problem, 
	which can be solved by water-filling~\cite{Cover1991IT}. 

	Let the eigenvalue decomposition (EVD) of ${\bf 
		\Sigma_h}$ be $\mb{U}_K{\bf \Lambda}_K\mb{U}_K^H$, where $K$ is the 
		rank of 
	$\bf \Sigma_h$, $\mb{U}_K = [\mb{u}_1, \mb{u}_2,\cdots,\mb{u}_K]$, and 
	${\bf 
		\Lambda}_K = {\rm diag}\{{\lambda}_1, 
	{\lambda}_2,\cdots,{\lambda}_K\}$ with 
	${\lambda}_1\ge\cdots\ge\lambda_K>0$. 
	The ideal observation matrix $\mb{W}^{\rm Ideal}$ is given 
	as 
	$\mb{W}^{\rm Ideal} 
	= 
	\mb{U}_K \mb{P}$. The power allocation matrix    
	$\mb{P}\in 
	\mathbb{R}^{K\times Q}$ is  
	expressed as $\mb{P} = 
	[\mb{P}_K, \mb{0}_{K\times (Q-K)}]$ with $\mb{P}_K = 
	{\rm diag}\{\sqrt{p_1}, \cdots, \sqrt{p_K}\}$. 
	The power $p_k$ allocated to 
	each eigenvector is calculated by the water-filling principle: 
	\begin{align}\label{eq:WF}
		p_k = (\beta - {\sigma^2}/{\lambda_k})^{+}, \forall 
		k \in \{1,2,\cdots, K\},
	\end{align}
	where {\color{black} 
		$(x)^+\overset{\Delta}{=}\max\{x, 
		0\}$}, 
	and 
	the water-level $\beta$ is properly selected to meet the power 
	constraint: $\sum_{k=1}^{K} 
	p_k = Q$.

	\section{Practical Observation Matrix Design}\label{sec:4}
	
	This section elaborates on the observation matrix design in practical 
	scenarios. Firstly, a greedy-method based principle is leveraged to solve 
	the MIM problem in \eqref{eq:MI}. Building on this principle, we propose an 
	ice-filling algorithm for designing the amplitude-and-phase controllable 
	observation matrices. Additionally, we conduct a comprehensive analysis on 
	the 
	relationship 
	between ice-filling and water-filling.  Subsequently, a MM-based algorithm 
	is proposed for 
	the design of phase-only controllable observation matrices. Last, the 
	choice of 
	prior kernel is also discussed.
	
	\subsection{Observation Matrix Design Using Greedy Method}
	\begin{figure}[!t]
		\centering
		\includegraphics[width=0.44\textwidth]{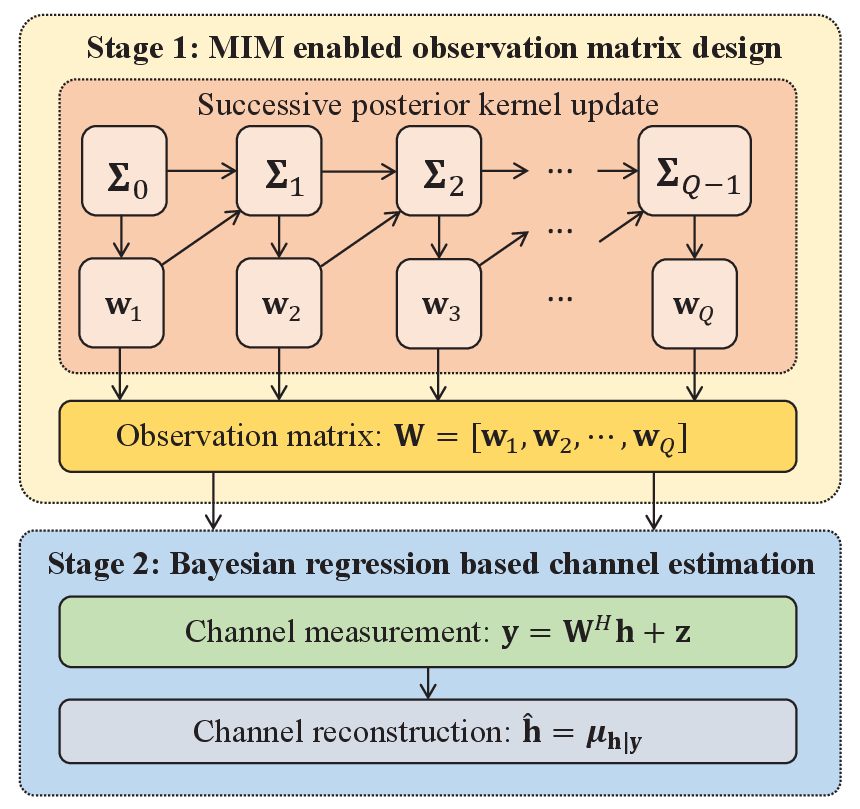}
		\vspace*{-0.5em}
		\caption{Framework of the proposed observation matrix design.}
		\vspace*{-1em}
		\label{img:framework}
	\end{figure}
	
	In practical uplink channel estimation where the noise vector $\mb{z}$ is 
	amplified by the observation vector $\mb{w}_q$, the 
	pilot-wise power constraint should be considered.  In this case, it is 
	intractable to find the optimal solution to \eqref{eq:MI}. To overcome 
	this challenge, we harness 
	a greedy method to obtain the observation vectors in a pilot-by-pilot 
	manner, as shown in Fig.~\ref{img:framework}. 
	Define ${\bf{W}}_t = [{\bf{w}}_1, 
	{\bf{w}}_2, \cdots, {\bf{w}}_t]$ as the observation 
	matrix for timeslots 
	$1\sim t$, where $t < Q$, and denote $\mb{y}_{t} 
	= \mb{W}_{t}^H\mb{h} + \mb{z}_{t}$ as the corresponding received signal. 
	Given the current observation 
	matrix ${\bf{W}}_t$, our greedy method aims to determine the next 
	observation 
	vector,
	$\mb{w}_{t+1}$, by maximizing the MI increment  
	from timeslot $t$ to $t+1$, i.e., $\max_{\mb{w}_{t+1}}
	I(\mb{y}_{t+1};\mb{h}) - 
	I(\mb{y}_t; \mb{h})$. 
	{\color{black} As proved in \cite{OptimalBF_Pi2012}, the MI increment is 
		expressed as
		\begin{align}\label{eq:MIIncrement}
			I(\mb{y}_{t+1};\mb{h}) - 
			I(\mb{y}_t; \mb{h}) = \log_2 \left(1 + 
			\frac{1}{\sigma^2}\mb{w}_{t+1}^H{\bm 
				\Sigma}_t\mb{w}_{t + 1}\right), 
		\end{align}
		where ${{\bf{\Sigma }}_t}$ represents the posterior kernel 
		of $\mb{h}$ given the observation $\mb{y}_t$.  
		Using the block matrix inverse, we can explicitly write 
		${\bf{\Sigma }}_t$ as 
		a function 
		of ${\bf{\Sigma }}_{t-1}$ and $\mb{w}_t$ (Detailed proof can be found 
		in 
		Appendix A): 
		\begin{align}\label{eq:lemma1}
			{\bf{\Sigma}}_t ={\bf{\Sigma}}_{t-1} - \frac{{\bf{\Sigma}}_{t-1}  
				{\bf{w}}_{t}
				{\bf{w}}_{t}^H{\bf{\Sigma}}_{t-1}}{{\bf{w}}_{t}^H{\bf{\Sigma}}_{t-1}{\bf{w}}_{t}
				+ \sigma^2}.
		\end{align}
		Accordingly, the optimal ${\bf{w}}_{t + 1}$ can be attained by solving 
		the 
		quadratic problem:
		\begin{align}\label{eq:RayleighEntropy}
			{\bf{w}}_{t + 1} = 
			\mathop{\arg\!\max}\limits_{{\bf{w}}\in\mathcal{W}} 
			{\bf{w}}^H 
			{\bf{\Sigma}}_t
			{\bf{w}}.
		\end{align}
	}
			%
	
	As a result, 
	equations \eqref{eq:MIIncrement} to \eqref{eq:RayleighEntropy} inspire us 
	to 
	propose 
	the framework in Fig. \ref{img:framework} for designing
	observation matrices. 
	It begins by setting ${\bm \Sigma}_0$ as the 
	prior kernel, ${\bm \Sigma}_{\bf h}$. Then, the first observation vector 
	$\mb{w}_1$ is obtained by solving problem \eqref{eq:RayleighEntropy}. In 
	subsequent timeslots $t \in 
	\{1,2,\cdots, Q-1\}$,  the posterior 
	kernels ${\bm \Sigma}_t$ are updated 
	from ${\bm \Sigma}_{t - 1}$ and  ${\bf w}_{t}$ using 
	\eqref{eq:lemma1}, while the observation 
	vectors ${\bf w}_{t + 1}$ are updated from ${\bm \Sigma}_t$ by solving 
	problem 
	\eqref{eq:RayleighEntropy}. 
	These 
	two updating steps are repeated until all $Q$ observation vectors are  
	generated for performing the ensuing Bayesian regression. 
	
	In the sequel, we will elaborate on solving problem 
	\eqref{eq:RayleighEntropy} under the 
	amplitude-and-phase controllable and phase-only controllable analog 
	combiners, 
	respectively.
	
	\subsection{Ice-filling for Amplitude-and-phase Controllable Observation 
	Matrix}
	\begin{algorithm}[!t]
		\caption{Ice-Filling-Based Observation Matrix Design} 
		\begin{algorithmic}[1]\label{alg:IF}
			\REQUIRE  
			Number of pilots $Q$, kernel ${\bm \Sigma}_{\bf h}$.
			\ENSURE 
			Designed observation matrix ${\bf W}$.
			\STATE Find the eigenvectors $[\mb{u}_1, \mb{u}_2, \cdots, 
			\mb{u}_K]$ 
			and the corresponding eigenvalues $[\lambda_1, \lambda_2, \cdots, 
			\lambda_K]$ of ${\bf 
				\Sigma}_0 = {\bf 
				\Sigma_h}$ 
			\STATE Initialize $[\lambda_1^0, \lambda_2^0, \cdots, 
			\lambda_K^0] = [\lambda_1, \lambda_2, \cdots, 
			\lambda_K]$
			\FOR{$t = 0, \cdots, Q-1$}
			\STATE $k_t = \arg\max_{k\in\{1,2,\cdots, K\}}\{\lambda_k^{t}\} 
			$ 
			\STATE Eigenvector-assignment: $\mb{w}_{t+1} = \mb{u}_{k_t}$
			\STATE Eigenvalue-update: $\lambda_{k_t}^{t+1}  = 
			\frac{\lambda_{k_t}^{t}\sigma^2}{\lambda_{k_t}^{t} + 
				\sigma^2}$
			\STATE Eigenvalue-preserve: $\lambda_{k}^{t+1} =  \lambda_{k}^{t}
			$ for $k \neq k_t$
			\ENDFOR
			\STATE Construct observation matrix: ${\bf W} = [\mb{w}_1, 
			\mb{w}_2, 
			\cdots, 
			\mb{w}_Q]$
			\RETURN Designed observation matrix ${\bf W}$ 
		\end{algorithmic}
	\end{algorithm}
	
	For amplitude-and-phase controllable observation matrices, problem 
	\eqref{eq:RayleighEntropy} is the well-known Rayleigh 
	quotient problem: 
	\begin{align}\label{eq:w1}
		{\bf{w}}_{t + 1} = \mathop{\arg\!\max}\limits_{\|{\bf w}\|_2^2 = 1} 
		{\bf{w}}^H 
		{\bf{\Sigma}}_t
		{\bf{w}}.
	\end{align}
	The principal 
	eigenvector 
	of ${\bf{\Sigma}}_t$ yields the optimal solution to $\mb{w}_{t+1}$, i.e., 
	${\bf{\Sigma}}_t{\bf{w}}_{t + 1} = 
	\lambda_{\max}  
	({\bf{\Sigma}}_t) {\bf{w}}_{t + 1}$. 
	By leveraging the updating strategy of ${\bm 
		\Sigma}_t$ described in \eqref{eq:lemma1} and the property of  
	eigenvector, we can prove  
	\textbf{Theorem~\ref{theorem:1}}, which lays the foundation for our  
	ice-filling 
	algorithm. 
	\begin{theorem}\label{theorem:1}
		If $\mb{w}_t$ is the principal eigenvector of ${\bm 
			\Sigma}_{t-1}$, then ${\bm 
			\Sigma}_t$ can be 
		rewritten as 
		\begin{align}\label{eq:thm1}
			{\bf {\Sigma}}_t = {\bf \Sigma}_{t-1} - \frac{\lambda_{\max}^2({\bf 
					\Sigma}_{t-1})}{\lambda_{\max}({\bf 
					\Sigma}_{t-1}) + \sigma^2}\mb{w}_t\mb{w}_t^H.
		\end{align}
	\end{theorem}
	
	\begin{IEEEproof}
		Applying the property of the principal eigenvector ${\bf 
			\Sigma}_{t-1}\mb{w}_t = \lambda_{\max}({\bm 
			\Sigma}_{t-1})\mb{w}_t$,  
		we 
		get 
		${\bf{w}}_{t}^H{\bf{\Sigma}}_{t-1}{\bf{w}}_{t}= \lambda_{\max}({\bf 
			\Sigma}_{t-1})$ and
		{\color{black} ${\bf{\Sigma}}_{t-1}  
			{\bf{w}}_{t}
			{\bf{w}}_{t}^H{\bf{\Sigma}}_{t-1} = \lambda_{\max}^2({\bf 
				\Sigma}_{t-1}) \mb{w}_t\mb{w}^H_t$}, which together with 
		\eqref{eq:lemma1}
		give rise to 
		(\ref{eq:thm1}). 
	\end{IEEEproof}
	\begin{remark}
		\emph{	Given that $\mb{w}_t$ is the principal eigenvector of ${\bm 
				\Sigma}_{t-1}$, 
			\textbf{Theorem 
				\ref{theorem:1}} implies that the posterior kernels, 
				${\bm\Sigma}_{t}$ and 
			${\bm\Sigma}_{t-1}$, share the identical eigenspace. 
			This fact immediately leads to the result that the eigenvectors, 
			$\mb{U}_K= [\mb{u}_1, \mb{u}_2, \cdots, 
			\mb{u}_K]$, of the prior kernel, 
			${\bm 
				\Sigma}_0 
			= 
			{\bm \Sigma}_{\mb{h}}$,  are inherited by all subsequent posterior 
			kernels, 
			${\bm \Sigma}_1, {\bm \Sigma}_2, \cdots,$ and ${\bm \Sigma}_{Q-1}$. 
			Thereby, we can draw the conclusion that all observation vectors, 
			$\{\mb{w}_1\}_{t=1}^Q$,  
			are selected from the eigenvectors of ${\bm \Sigma}_{\mb{h}}$.  }
	\end{remark}
	\begin{remark}	\color{black}
		\emph{Consider the eigenvalues of ${\bf \Sigma}_{t-1}$ and ${\bf 
				\Sigma}_{t}$. We denote the EVD of ${\bf 
				\Sigma}_{t-1}$ as ${\bf 
				\Sigma}_{t-1} = \mb{U}_K{\rm diag}\{\lambda_1^{t-1}, 
				\lambda_2^{t-1}, 
			\cdots, 
			\lambda_K^{t-1}\}\mb{U}_K^H$, where  $\{\lambda_k^{t-1}\}_{k = 
			1}^K$ refers 
			to the eigenvalues of $\{\mb{u}_k\}_{k = 1}^K$, respectively. We 
			use 
			$k_{t-1}$ to index the 
			principal eigenvalue of ${\bf \Sigma}_{t-1}$. Accordingly, the 
			equations 
			$\mb{w}_t = \mb{u}_{k_{t-1}}$ and $\lambda_{\max}({\bf 
			\Sigma}_{t-1}) = 
			\lambda_{k_{t-1}}^{t-1}$ hold, and \textbf{Theorem~\ref{theorem:1}} 
			can be 
			rewritten as:
			\begin{align}\label{eq:remark2}
				{\bf {\Sigma}}_t &= \mb{U}_K{\rm diag}\{\lambda_1^{t-1}, 
				\cdots, 
				\lambda_{k_{t-1}}^{t-1}, \cdots, 
				\lambda_K^{t-1}\}\mb{U}_K^H  \notag \\ 
				&\quad\quad\quad\quad\quad\quad- 
				\frac{(\lambda_{k_{t-1}}^{t-1})^2}{\lambda_{k_{t-1}}^{t-1} + 
					\sigma^2}\mb{u}_{k_{t-1}}\mb{u}_{k_{t-1}}^H \\
				& = \mb{U}_K{\rm diag}\left\{\lambda_1^{t-1}, \cdots, 
				\frac{\lambda_{k_{t-1}}^{t-1}\sigma^2}{\lambda_{k_{t-1}}^{t-1} 
				+ 
					\sigma^2},
				\cdots, 
				\lambda_K^{t-1}\right\}\mb{U}_K^H. \notag
			\end{align}
			Equation \eqref{eq:remark2} reveals that only the 
			principal 
			eigenvalue, $\lambda_{k_{t - 1}}^{t-1}$, of ${\bm \Sigma}_{t 
				- 1}$ is squeezed to an eigenvalue of 
			${\bm \Sigma}_{t}$ given by 
			$\frac{\lambda_{k_{t-1}}^{t-1}\sigma^2}{\lambda_{k_{t-1}}^{t-1} 
				+ 
				\sigma^2}$,  while the remaining 
			eigenvalues, $\lambda_{k}^{t - 1}, \forall 
			k\in\{1,2,\cdots,K\}/\{k_{t-1}\}$,  keep unchanged.  
		}
	\end{remark}
	
	Motivated by \textbf{Remark 1} and \textbf{Remark 2}, our ice-filling 
	algorithm is intrinsically an \emph{eigenvector-assignment-process}, 
	as presented in \textbf{Algorithm~\ref{alg:IF}}. To elaborate, 
	the eigenvalues $\{\lambda_1^t, \lambda_2^t, \cdots, \lambda_K^t\}$ of the 
	posterior kernel
	${\bm \Sigma}_t$ are initialized as the eigenvalues of the prior 
	kernel ${\bm 
		\Sigma}_{\bf h}$ at $t = 0$ in Step~2. In each 
	timeslot, we find the largest eigenvalue 
	from  $\{\lambda_1^{t},\lambda_2^{t},\cdots,\lambda_K^{t}\}$ and index it 
	by 
	$k_t$ in Step~4. 
	Then,  the corresponding eigenvector $\mb{u}_{k_t}$ of ${\bm\Sigma_h}$, 
	which 
	is equivalent to 
	the principal eigenvector of ${\bm\Sigma}_{t}$, is assigned to 
	$\mb{w}_{t+1}$ 
	in 
	Step 5.  Finally, according to \textbf{Remark 2}, the selected 
	eigenvalue $\lambda_{k_t}^{t}$ is squeezed to 
	$\frac{\lambda_{k_t}^{t}\sigma^2}{\lambda_{k_t}^{t} + 
		\sigma^2}$,  while the other eigenvalues of ${\bm \Sigma}_{t}$ are 
	preserved,  
	to attain the 
	eigenvalues 
	of the next kernel ${\bm \Sigma}_{t+1}$. These steps are 
	repeatedly executed until all observation vectors are generated, which 
	completes the ice-filling algorithm.

	\subsection{Ice-Filling Versus Water-Filling} 
	\begin{figure*}[!t]
		\centering
		\includegraphics[width=1\textwidth]{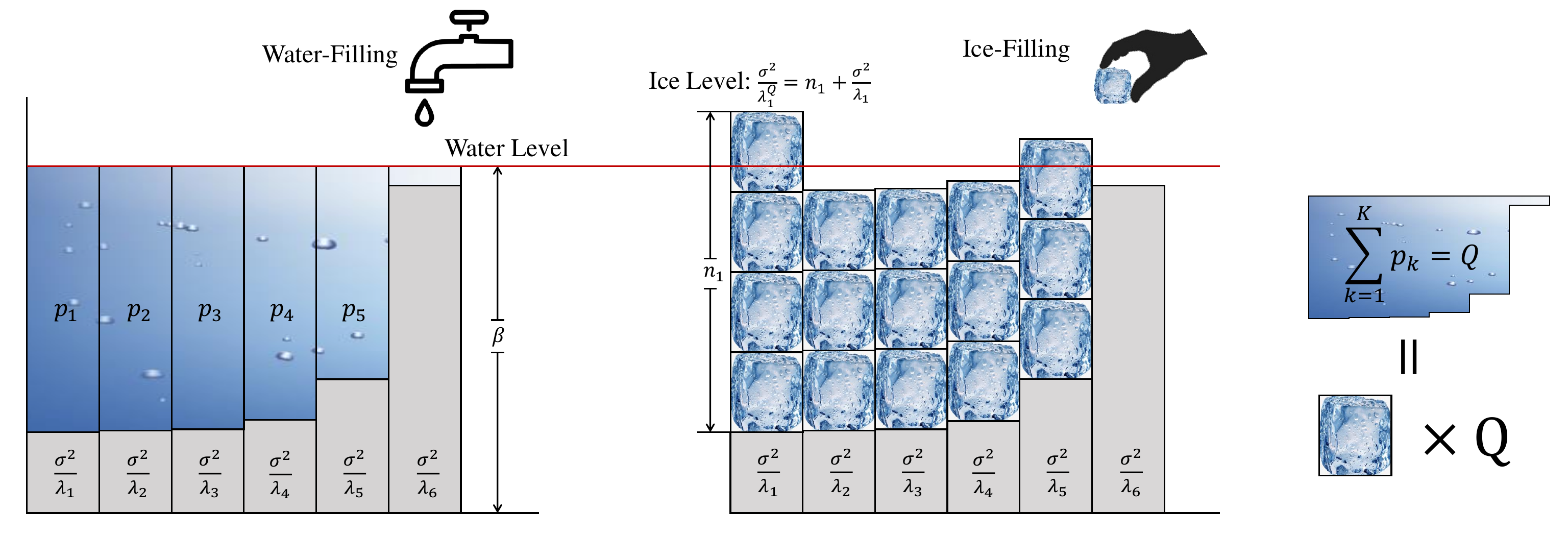}
		\vspace*{-2em}
		\caption{Comparison between water-filling and ice-filling. The rank of 
			the 
			prior kernel, the number of antennas, and the total pilot 
			length are set as $K = 6$, $M = 
			128$, and $Q = 16$, respectively.  The eigenvalues of 
			the prior kernel, ${\bf \Sigma}_{\mb{h}}$, are in a descending 
			order, i.e., $\lambda_1 > 
			\lambda_2 > \lambda_3 > \lambda_4 > \lambda_5 > \lambda_6$.
		}
		\vspace*{-1em}
		\label{img:icefilling}
	\end{figure*}
	In this subsection, we put forward a more insightful interpretation to the 
	ice-filling algorithm that 
	establishes its connection with the  water-filling principle. Generally 
	speaking, the 
	proposed ice-filling algorithm can be viewed as a 
	quantization of the water-filling principle. 
	Specifically, it is clear that the observation matrices generated by both 
	the water-filling 
	and the ice-filling algorithms fall within the eigenspace of ${\bm 
		\Sigma}_{\bf 
		h}$. The major distinction is attributed to their ``power allocation 
	strategies". 
	The water-filling 
	principle allocates $Q$ units of power to the $K$ eigenvectors using 
	\eqref{eq:WF}.  The ice-filling algorithm  transforms this power-allocation 
	process into an \emph{eigenvector-assignment process}, where   
	the $K$ eigenvectors are repeatably assigned to the observation vectors in 
	$Q$ 
	timeslots. 
	
	To be more precise, recall that $p_k$  in \eqref{eq:WF}
	denotes the optimal power allocated to the $k$-th eigenvector, satisfying 
	$\sum_{k=1}^K p_k = Q$. For a clear comparison, 
	we introduce the 
	definition of 
	``pilot reuse frequency" as follows. 
	\begin{definition}\label{def:1}
		The pilot reuse frequency $n_k^t \in \mathbb{Z}^{+}$, with 
		$k\in\{1,2,\cdots,K\}$ and $t\in\{1,2,\cdots, Q\}$, is defined as the 
		number of 
		times that the $k$-th eigenvector, $\mb{u}_k$, of the prior kernel is  
		selected as the observation vector by the ice-filling algorithm 
		during 
		timeslots $1\sim t$. The pilot reuse frequencies satisfy $\sum_{k=1}^K 
		n_k^t = t$. For 
		ease 
		of 
		expression, we  define $n_k := n_k^Q$ as the pilot reuse frequency 
		during 
		the 
		total $Q$ timeslots. 
	\end{definition}
	
	Comparing the definitions of $p_k$ and $n_k$, we can intuitively interpret 
	the 
	$n_k$-timeslot reuse of the observation vector $\mb{u}_k$ for pilot 
	transmission 
	as allocating $n_k$ units of power to 
	$\mb{u}_k$.
	More importantly, \textbf{Definition \ref{def:1}} enables us to derive the 
	analytical expressions of the 
	eigenvalues $\{\lambda_k^t\}_{k=1}^K$ in \textbf{Algorithm \ref{alg:IF}}. 
	Note that the eigenvalue update rule of the ice-filling algorithm can 
	be 
	rewritten as a recursive formula:
	\begin{align} \label{eq:recur}
		\lambda_{{k_t}}^{t + 1} = 
		\frac{\lambda_{k_t}^t\sigma^2}{\lambda_{k_t}^t + 
			\sigma^2}	
		\Leftrightarrow \frac{\sigma^2}{\lambda_{k_t}^{t + 1}} = 1 + 
		\frac{\sigma^2}{\lambda_{k_t}^{t}}, 
	\end{align}
	where $\lambda_{k_t}^{t}$ is the largest eigenvalue from 
	$\{\lambda_1^{t},\lambda_2^{t},\cdots,\lambda_K^{t}\}$. This recursive 
	formula indicates that whenever the $k$-th eigenvector is selected as an 
	observation vector, the value of  $\frac{\sigma^2}{\lambda_k^{t}}$ 
	increases by 
	1. By further considering the definition of the pilot reuse frequency 
	$n_k^t$, 
	that 
	the $k$-th eigenvector is 
	selected by $n_k^t$ times during timeslots $1\sim t$, we can naturally 
	obtain 
	the relationship between $n_k^t$ and $\frac{\sigma^2}{\lambda_k^{t}}$:
	\begin{align}\label{eq:nk}
		\frac{\sigma^2}{\lambda_k^{t}} = n_k^t + \frac{\sigma^2}{\lambda_k} 
		\Leftrightarrow n_k^t = \frac{\sigma^2}{\lambda_k^{t}} - 
		\frac{\sigma^2}{\lambda_k}, \:\:k\in\{1,\cdots,K\}.
	\end{align}
The comparison between equations \eqref{eq:nk} and \eqref{eq:WF} allows us to 
interpret the ice-filling algorithm as a 
quantization of the water-filling algorithm, as 
elaborated below.

\textbf{Interpretation of ice-filling}: 
As depicted in Fig. \ref{img:icefilling}, the water-filling principle 
allocates 
$Q$ 
units of water (power) to a vessel with $K$ channels, each having a unique 
base 
level 
$\frac{\sigma^2}{\lambda_k}$,  $k\in\{1,2,\cdots, K\}$. By controlling 
the \emph{uniform} 
water level $\beta$, the 
optimal power $p_k$ is determined by the gap between the base level and the 
water level. 
In contrast, the ice-filling algorithm transforms the total $Q$ units 
of
water into $Q$ ice blocks, each containing one unit of 
power and being used for one 
pilot transmission.  
Our algorithm starts from an empty vessel and  fills one ice block onto 
the channel 
having the deepest base surface $\frac{\sigma^2}{\lambda_{k_0}}$. This 
operation is equivalent to finding 
the 
largest eigenvalue $\lambda_{k_0}$ from $\{\lambda_1, \cdots, 
\lambda_K\}$. Then,  the eigenvector 
$\mb{u}_{k_0}$ is 
assigned 
to the first observation vector $\mb{w}_1$, and ice 
level 
of this channel increases from  $\frac{\sigma^2}{\lambda_{k_0}}$ 
to $\frac{\sigma^2}{\lambda_{k_0}^1} = \frac{\sigma^2}{\lambda_{k_0}} + 1$. 
In the subsequent time slots, the remaining $Q -1$ ice blocks are 
filled onto the channels locating at the deepest base surfaces or ice 
levels, 
indexed by $k_t = \arg\min_k\{\frac{\sigma^2}{\lambda_k^{t}}\}_{k=1}^K$, 
one by 
one. 
The 
corresponding eigenvectors $\mb{u}_{k_t}$ are used for pilot transmission,  
and 
the ice levels increase according to \eqref{eq:nk}.
Consequently, the final pilot reuse frequencies $\{n_k\}_{k=1}^K$ are 
determined by 
the 
number of ice blocks on top of each channel, and the ice levels are given 
by 	
$\{\frac{\sigma^2}{\lambda_k^{Q}}\}_{k=1}^K = \{n_k + 
\frac{\sigma^2}{\lambda_k}\}_{k=1}^K$.

The comparison between the ice-filling and water-filling in Fig. 
\ref{img:icefilling} indicates that the continuous powers 
$\{p_k\}_{k=1}^K$ are quantized into the discrete pilot reuse 
frequencies $\{n_k\}_{k=1}^K$. This quantization   
results in the non-uniform ice levels 
$\{\frac{\sigma^2}{\lambda_{k}^Q}\}_{k=1}^K$, as 
opposed 
to the 
uniform water level $\beta$. 
To see the quantization nature of ice-filling more clearly, we prove 
\textbf{Theorem \ref{theorem:2}} to upper bound the quantization error 
between $p_k$ and 
$n_k$.

\begin{theorem}\label{theorem:2}
	Considering the $k$-th pilot reuse frequency, $n_k$, and the $k$-th  
	optimal power, $p_k$, we have
	\begin{align}
		|n_k - p_k| < 1.
	\end{align}
\end{theorem}
\begin{IEEEproof}
	See Appendix B. 
\end{IEEEproof}
\begin{remark}
	\emph{\textbf{Theorem \ref{theorem:2}} rigorously proves that the deviation 
		of 
		$n_k$ from $p_k$ is less than 1, rendering the pilot reuse 
		frequency a good approximation of the optimal power allocation. 
		Particularly, when 
		the 
		total 
		pilot length $Q$ is large, the relative error between $n_k$ and $p_k$ 
		tends 
		to 
		zero because
		\begin{align}
			\frac{|n_k - p_k|}{p_k} < \frac{1}{p_k} \overset{Q \rightarrow 
				+\infty}{=} 0.
		\end{align}
		Thereafter, the proposed ice-filling algorithm features a discrete 
		approximation to the ideal water-filling algorithm with the relative
		quantization 
		error decaying rapidly.} \end{remark}

\subsection{Majorization Minimization for Phase-Only Controllable Combiner}
Turn now to the phase-only controllable analog combiners where the 
amplitude of each element of $\bf W$ is fixed as a constant. The observation 
vector at the $(t+1)$-th timeslot 
can 
be obtained 
by
\begin{equation}\label{eq:Phase_only}
	{{\bf{w}}_{t + 1}} = \mathop {\arg\!\max }\limits_{\left| {\bf{w}} \right| 
	= \frac{1}{{\sqrt M }}{{\bf{1}}_M}} {{\bf{w}}^H}{{\bf{\Sigma }}_t}{\bf{w}}.
\end{equation}
Due to the unit modulus constraint, the problem (\ref{eq:Phase_only}) is a 
non-convex programming. To cope with this problem, we propose a 
MM-based observation matrix design, as presented in \textbf{Algorithm 
	\ref{alg:mm}}. 
Its key idea is to reformulate the 
non-convex problem as a series of more tractable approximate subproblems, which 
can be 
solved in an alternating manner \cite{SunYing'17'TSP}. Specifically, the
proposed MM-based method is illustrated as follows. 

\begin{algorithm}[!t]
	\caption{MM-Based Observation Matrix Design} 
	\begin{algorithmic}[1]\label{alg:mm}
		\REQUIRE  
		Number of pilots $Q$, kernel ${\bm \Sigma}_{\bf h}$.
		\ENSURE 
		Designed observation matrix ${\bf W}$.
		\STATE Initialization: ${\bf W}=\varnothing$, ${{\bf{\Sigma }}_0}={\bm 
		\Sigma}_{\bf h}$.
		\FOR{$t=0,\cdots,Q-1$}
		\STATE Initialize ${\bf w}_{t+1}$ randomly
		\STATE Update: ${\bf{X}} = \mathsf{Tr}\left( {{\lambda _{\max }}\left( 
			{{{\bf{\Sigma }}_t}} \right){{\bf{I}}_M} - {{\bf{\Sigma }}_t}} 
		\right){{\bf{I}}_M}$
		\WHILE{no convergence of ${{\bf{w}}^H_{t+1}}{{\bf{\Sigma 
		}}_t}{\bf{w}}_{t+1}$}
		\STATE Update: ${\bf v}={\bf w}_{t+1}$
		\STATE Update the $t+1$-th observation vector: ${\bf{w}}_{t+1} = 
		\frac{1}{{\sqrt M }}{\exp\left(j\angle \left( {{\bf{X}} - {\lambda 
					_{\max }}\left( {{{\bf{\Sigma }}_t}} \right){{\bf{I}}_M} + 
					{{\bf{\Sigma 
					}}_t}} \right){\bf{v}}\right)}$
		\ENDWHILE
		\STATE Update kernel: ${\bf{\Sigma}}_{t+1} ={\bf{\Sigma}}_{t} - 
		\frac{{\bf{\Sigma}}_{t}  
			{\bf{w}}_{t+1}
			{\bf{w}}_{t+1}^H{\bf{\Sigma}}_{t}}{{\bf{w}}_{t+1}^H{\bf{\Sigma}}_{t}{\bf{w}}_{t+1}
			+ \sigma^2}$
		\ENDFOR
		\STATE Construct observation matrix: ${\bf W} = [\mb{w}_1, \mb{w}_2, 
		\cdots, \mb{w}_Q]$
		\RETURN Designed observation matrix ${\bf W}$ 
	\end{algorithmic}
\end{algorithm}

To begin with, given that adding a constant 
to 
the objective function does not affect the optimal solution of 
(\ref{eq:Phase_only}), problem (\ref{eq:Phase_only}) is equivalent to
\begin{align}\label{eq:Phase_only_2}
	\notag
	{{\bf{w}}_{t + 1}} \overset{(a)}{=} &\mathop {\arg\!\max }\limits_{\left| 
		{\bf{w}} \right| = \frac{1}{{\sqrt M }}{{\bf{1}}_M}} 
		{{\bf{w}}^H}{{\bf{\Sigma 
		}}_t}{\bf{w}} - {\lambda _{\max }}\left( {{{\bf{\Sigma }}_t}} 
	\right){{\bf{w}}^H}{\bf{w}} \\
	=& \mathop {\arg\!\min }\limits_{\left| {\bf{w}} \right| = \frac{1}{{\sqrt 
	M 
		}}{{\bf{1}}_M}} {{\bf{w}}^H}\left( {{\lambda _{\max }}\left( 
		{{{\bf{\Sigma 
				}}_t}} \right){{\bf{I}}_M} - {{\bf{\Sigma }}_t}} 
				\right){\bf{w}},
\end{align}
where (a) holds because $\mb{w}^H\mb{w} = 1$.
Then, we consider to solve problem (\ref{eq:Phase_only_2}) by iteratively 
minimizing the 
upper-bound of its objective function. To this end, we introduce an auxiliary 
variable ${\bf v}\in{\mathbb C}^{M\times 1}$ to obtain an  
upper-bound function ${\bar f}({\bf w},{\bf v})$, given by 
\cite{Junxiao'16'TSP}:
\begin{align}\label{eq:upper-bound}
	\notag
	{{\bf{w}}^H}&\left( {{\lambda _{\max }}\left( {{{\bf{\Sigma }}_t}} 
	\right){{\bf{I}}_M} - {{\bf{\Sigma }}_t}} \right){\bf{w}} \le {\bar f}({\bf 
	w},{\bf v}) = \\ & \notag {{\bf{w}}^H}{\bf{Xw}} - 2\Re \left\{ 
	{{{\bf{w}}^H}\left( {{\bf{X}} - {\lambda _{\max }}\left( {{{\bf{\Sigma 
	}}_t}} \right){{\bf{I}}_M} + {{\bf{\Sigma }}_t}} \right){\bf{v}}} \right\} 
	+ \\ & {{\bf{v}}^H}\left( {{\bf{X}} - {\lambda _{\max }}\left( 
	{{{\bf{\Sigma }}_t}} \right){{\bf{I}}_M} + {{\bf{\Sigma }}_t}} 
	\right){\bf{v}},
\end{align}
where ${\bf{X}} = \mathsf{Tr}\left( {{\lambda _{\max }}\left( {{{\bf{\Sigma 
			}}_t}} \right){{\bf{I}}_M} - {{\bf{\Sigma }}_t}} 
			\right){{\bf{I}}_M}$ and the 
equality holds when ${\bf v}={\bf w}$. In this way, ${\bf w}$ and ${\bf v}$ in 
the upper-bound function ${\bar f}({\bf w},{\bf v})$ can be alternatively 
optimized to approach a sub-optimal solution to problem (\ref{eq:Phase_only_2}) 
\cite{SunYing'17'TSP}. In particular, for a given ${\bf w}$, ${\bf v}$ can be 
updated by ${\bf v}={\bf w}$. For a given ${\bf v}$, by utilizing 
${{\bf{w}}^H}{\bf{X}}{\bf{w}}=\mathsf{Tr}\left( {{\lambda _{\max }}\left( 
	{{{\bf{\Sigma }}_t}} \right){{\bf{I}}_M} - {{\bf{\Sigma }}_t}} \right)$ and 
removing the unrelated part in ${\bar f}({\bf w},{\bf v})$ in 
(\ref{eq:upper-bound}), the subproblem of updating $\bf w$ can be rewritten as
\begin{align}\label{eq:w_t+1}
	\notag
	&{{\bf{w}}}_{t+1} = \\ & \mathop {\arg\!\max }\limits_{\left| {\bf{w}} 
	\right| = \frac{1}{{\sqrt M }}{{\bf{1}}_M}} \Re \left\{ {{{\bf{w}}^H}\left( 
	{{\bf{X}} - {\lambda _{\max }}\left( {{{\bf{\Sigma }}_t}} 
	\right){{\bf{I}}_M} + {{\bf{\Sigma }}_t}} \right){\bf{v}}} \right\}.
\end{align}
It can be easily proved that the optimal solution to (\ref{eq:w_t+1}) is given 
by
${{\bf{w}}}_{t+1} = \frac{1}{{\sqrt M }}{\exp\left(j\angle \left( {{\bf{X}} - 
		{\lambda _{\max }}\left( {{{\bf{\Sigma }}_t}} \right){{\bf{I}}_M} + 
		{{\bf{\Sigma }}_t}} \right){\bf{v}}\right)}$.
This solution, associated with 
the updating approach of the posterior kernel in \eqref{eq:lemma1}, 
completes the 
algorithm. It 
is worth noting that, since the updates of $\bf v$ and ${\bf w}_{t+1}$ both 
monotonously increase the objective function, the convergence of 
\textbf{Algorithm \ref{alg:mm}} is naturally guaranteed.

\subsection{Kernel Selection in Non-Ideal Case}\label{subsec:4:D}
In some non-ideal cases where the perfect prior kernel, 
${\bf\Sigma}_{\bf h}$, is difficult to acquire, the initialization states of  
\textbf{Algorithms \ref{alg:IF} and \ref{alg:mm}} may pose a challenge. To deal 
with this issue, we suggest two methods to obtain imperfect prior 
kernels. 
\subsubsection{Statistical kernel}
Firstly,  the prior kernel can be selected as the statistical 
covariance of historical estimated channels, i.e., ${\bf{\hat \Sigma }}_{{\bf 
		h}} = \mathsf{E}( {{\bf{\hat h}}{{{\bf{\hat h}}}^H}} )$. Due to the 
error of channel estimation, the historical channels ${\bf{\hat  
		h}}$ deviate from the true channels ${\bf{
		h}}$. We use a Gaussian noise to model this 
deviation, and get  
$
{\bf{\hat h}} = {\bf{h}} + 
{{\bf{z}}_{\bf{h}}}
$,
wherein ${{\bf{z}}_{\bf{h}}}\sim{\cal CN}\left( 
{{{\bf{0}}_M},\sigma _{\bf{h}}^2{{\bf{I}}_M}} \right)$ 
denotes the historical channel estimation error. For simplicity, we name 
$\sigma _{\bf{h}}^2$ as the {\it kernel 
	estimation error}. Thereby, the statistical kernel is modeled as
\begin{align}\label{eq:statis}
	{\bf{\hat \Sigma }}_{{\bf 
			h}} = {\bf{ \Sigma }}_{{\bf 
			h}} + \sigma^2_{\mb{h}}\mb{I}_M.
\end{align}

\subsubsection{Artificial kernels}
Another choice is to employ artificially designed kernels to mimic the perfect 
kernel. To this end, the artificial kernels 
should assign high similarity to nearby antennas while diminishing the 
correlation rapidly with inter-antenna distance. For example, here we consider 
a general uniform planer array (UPA) equipped with $M$ antennas with an antenna 
spacing of $d$. 
The numbers of its horizontal antennas and vertical antennas are $M_x$ and 
$M_y$, respectively. Striking a balance between complexity and practicality, we 
recommend two artificial kernels as follows.

\begin{itemize}
	\item \textbf{Exponential kernel}: The exponential kernel ${\bm 
		\Sigma}_{\rm exp} $, 
	is the most popular selection in Bayesian estimation. Let ${{\bf{m}}_x} = 
	\left[ { - \frac{{{M_x} - 1}}{2}, - \frac{{{M_x} - 3}}{2}, \cdots 
		,\frac{{{M_x} - 1}}{2}} \right]^T$ and ${{\bf{m}}_y} = \left[ { - 
		\frac{{{M_y} - 1}}{2}, - \frac{{{M_y} - 3}}{2}, \cdots ,\frac{{{M_y} - 
				1}}{2}} \right]^T$. Then, the exponential kernels for the two 
				dimensions of 
	UPA, ${{\bf{\Sigma }}_{{\rm exp},x}}\in{\mathbb C}^{M_x\times M_x}$ and 
	${{\bf{\Sigma }}_{{\rm exp},y}}\in{\mathbb C}^{M_y\times M_y}$, can be 
	respectively expressed as
	\begin{align}
		{{\bf{\Sigma }}_{{\rm exp},x}} \!=\! \exp \left( { \!- {\eta_1 
				^2}{\frac{4\pi^2d^2}{\lambda^2}}{{\left| {{{\bf{1}}^T_{{M_x}}} 
						\otimes 
						{{\bf{m}}_x} \!-\! {\bf{m}}_x^T \otimes 
						{{\bf{1}}_{{M_x}}}} \right|}^{ 
					\odot 2}}} \right),\\
		{{\bf{\Sigma }}_{{\rm exp},y}} \!=\! \exp \left( { \!- {\eta_1 
				^2}{\frac{ 4\pi^2d^2}{\lambda^2}}{{\left| {{{\bf{1}}^T_{{M_y}}} 
						\otimes 
						{{\bf{m}}_y} \!-\! {\bf{m}}_y^T \otimes 
						{{\bf{1}}_{{M_y}}}} \right|}^{ 
					\odot 2}}} \right),
	\end{align}
	where $\eta_1$ is an adjustable hyper-parameter and ${\bf X}^{\odot 2}$ 
	denotes the element-wise product of two matrices $\bf X$. Then, the overall 
	kernel can be written 
	as
	\begin{equation}\label{eq:exp}
		{\bm \Sigma}_{\rm exp} = 	{{\bf{\Sigma }}_{{\rm exp},x}} \otimes 
		{{\bf{\Sigma }}_{{\rm exp},y}}.
	\end{equation}
	{\color{black} 	The exponential kernel, ${\bm \Sigma}_{\rm exp}$, is 
		a heuristic kernel. It exhibits the highest correlation at the diagonal 
		elements, with the channel correlation diminishing exponentially as the 
		antenna spacing increases. Compared to other kernels, the exponential 
		kernel shows reduced sensitivity to outliers, making it suitable for 
		reconstructing channels that lack obvious regularity.}

	\item \textbf{Bessel kernel}:
	{\color{black} The Bessel kernel, denoted as ${\bm \Sigma}_{\rm bes}$, is 
		well-suited for capturing and modeling complex-valued data exhibiting 
		oscillatory  patterns. It characterizes the covariance of array 
		steering vector when the user distributes uniformly in the spatial 
		domain. 
		For the considered UPA,  the Bessel kernel can be obtained as ${\bm 
			\Sigma}_{\rm bes} = \mathsf{E}(\mb{a}(\theta, \phi)\mb{a}^H(\theta, 
			\phi))$, 
		where 
		$\mb{a}(\theta, \phi)$ is the array 
		steering vector of UPA, and $\theta \sim \mathcal{U}(-\pi/2, \pi/2)$ 
		and $\phi\sim \mathcal{U}(-\pi/2, \pi/2)$ denote the 
		angle-of-arrivals.} 
	By calculating expectation, the Bessel kernel is expressed as
	\begin{align}
		{{\bf{\Sigma }}_{{\rm{bes}},x}} &= {J_0}\left( {\eta_2 
			\frac{2\pi d}{\lambda}\left| {{\bf{1}}_{{M_x}}^T \otimes 
			{{\bf{m}}_x} - 
				{\bf{m}}_x^T \otimes {{\bf{1}}_{{M_x}}}} \right|} \right),\\
		{{\bf{\Sigma }}_{{\rm{bes}},y}} &= {J_0}\left( {\eta_2 
			\frac{2\pi d}{\lambda}\left| {{\bf{1}}_{{M_y}}^T \otimes 
			{{\bf{m}}_y} - 
				{\bf{m}}_y^T \otimes {{\bf{1}}_{{M_y}}}} \right|} \right), 
	\end{align}
	where $J_0$ is the zero-order Bessel function of the first kind and 
	$\eta_2$ is 
	a hyper-parameter. Thus, the 
	overall kernel can be written as
	\begin{equation}\label{eq:bes}
		{{\bf{\Sigma }}_{{\rm{bes}}}} = {{\bf{\Sigma }}_{{\rm{bes}},x}} \otimes 
		{{\bf{\Sigma }}_{{\rm{bes}},y}}.
	\end{equation}
\end{itemize}

\section{Performance Analysis}\label{sec:5}
To evaluate the performance of the proposed MIM-based observation matrix 
design, 
this section provides comprehensive MSE analyses under different algorithmic 
conditions.

\subsection{Estimation Accuracy Under Perfect Kernel}
When the perfect prior kernel 
${{\bf{\Sigma 
}}}_{\mb{h}}$ is available,  the MSEs achieved by the proposed algorithms 
can be determined by the trace 
of their 
posterior kernels, i.e., 
\begin{align}
	\mathsf{E}(\|{\bm{\mu}_{\mb{h}|\mb{y}}} - 
	\mb{h}\|^2_2) =  \mathsf{Tr}\left({{\bf{\Sigma }}}_{\mb{h}|\mb{y}}\right). 
\end{align}
The subsequent two lemmas give out the close-form MSEs and their asymptotic 
expressions of the water-filling-based and ice-filling-based channel 
estimators, 
respectively.

\begin{lemma}\label{lemma:2}
	If the perfect kernel ${{\bf{\Sigma 
	}}}_{\mb{h}}$ is adopted, the MSE 
	$\delta_{\rm wf}$ of the 
	water-filling 
	algorithm is given by 
	\begin{align}\label{eq:lemma2}
		\delta_{\rm wf} =  \sum_{k = 1}^{K} \frac{\lambda_k \sigma^2}{p_k 
			\lambda_k + \sigma^2} \overset{Q\rightarrow +\infty}{=} 
		\mathcal{O}(K^2Q^{-1}). 
	\end{align}
\end{lemma}
\begin{IEEEproof}
	See Appendix C.
\end{IEEEproof}


\begin{lemma}\label{lemma:3}
	If the perfect kernel ${{\bf{\Sigma 
	}}}_{\mb{h}}$ is adopted, the MSE $\delta_{\rm if}$ of the 
	ice-filling 
	algorithm is given by
	\begin{align}\label{eq:lemma3}
		\delta_{\rm if} =  \sum_{k = 1}^{K} \frac{\lambda_k 
			\sigma^2}{n_k 
			\lambda_k + \sigma^2} \overset{Q\rightarrow +\infty}{=} 
		\mathcal{O}(K^2Q^{-1}).
	\end{align}
\end{lemma}
\begin{IEEEproof}
	See Appendix D.
\end{IEEEproof}
We can gain two insights from \textbf{Lemma~\ref{lemma:2}} and 
\textbf{Lemma~\ref{lemma:3}}.
Firstly, the MSEs achieved by the water-filling and ice-filling methods have 
identical asymptotic expressions,  both of which 
decay at the rate of 
$Q^{-1}$. The only 
difference is owing to the quantization of  
the continuous 
powers $\{p_k\}$ in \eqref{eq:lemma2} to the pilot reuse 
frequencies $\{n_k\}$ in 
\eqref{eq:lemma3}.  
Next, the MSEs decay quadratically as the rank of the channel kernel, $K$, 
decreases. This finding demonstrates the superiority of DASs in channel 
estimation. A denser antenna deployment increases the correlation between 
inter-antenna channels, decreases the rank of the channel kernel, and thereby 
improves the channel estimation accuracy.

Taking into account the quantization nature 
revealed in \textbf{Theorem \ref{theorem:2}}, the 
asymptotic achievability bound of the MSE of the ice-filling 
algorithm can be evaluated by the following theorem. 

%

\begin{theorem}\label{thm:3}
	As $Q\rightarrow +\infty$, the asymptotic MSE difference, $|\delta_{\rm 
		wf} - 
	\delta_{\rm if}|$,  is given by 
	\begin{align}
		|\delta_{\rm wf} -
		\delta_{\rm if}| = \mathcal{O}(K^3Q^{-2}). 
	\end{align}
\end{theorem}
\begin{IEEEproof}
	See Appendix E. 
\end{IEEEproof}
\begin{remark}
	\emph{\textbf{Theorem~\ref{thm:3}} guarantees that the MSE gap between the 
		ice-filling and the ideal water-filling algorithms 
		decays at a rate of $K^3Q^{-2}$, which demonstrates the near-optimality
		of the proposed ice-filling channel estimator when $Q \rightarrow 
		+\infty$ 
		or $K\rightarrow 0$. 
	}
\end{remark}

To validate the superiority of the proposed design, it is of great interest to 
evaluate the MSE achieved by the Bayesian regression \eqref{eqn:bayesian} using 
randomly generated observation matrices. For amplitude-and-phase controllable 
cases, we assume that all elements of $\mb{W}$ are independently generated from 
a Gaussian distribution $\mathcal{CN}(0, 1/M)$. For phase-only controllable 
cases, the angles of all elements of $\bf W$ are randomly selected from 
$[-\pi,+\pi]$. Then, we derive the asymptotic MSE of a random observation 
matrix in \textbf{Lemma~\ref{lemma:rand_MM}}.

\begin{lemma}\label{lemma:rand_MM}
	Assume that the perfect kernel ${{\bf{\Sigma}}}_{\mb{h}}$ is adopted and 
	$Q$ is sufficiently large. When $\bf W$ is randomly generated, the MSE 
	$\delta_{\rm rnd}$ achieved in both phase-and-amplitude  
	and phase-only controllable 
	cases can be approximated by
	\begin{align}\label{wqn:MM_MSE}
		\delta_{\rm rnd}  \approx	\sum\limits_{k = 1}^K {{{{\lambda 
						_k}{\sigma ^2}} \over {{Q \over M}{\lambda _k} + 
						{\sigma ^2}}}}  \overset{Q\rightarrow +\infty}{=}  
		\mathcal{O}(KMQ^{-1}),
	\end{align}
	where the approximate equality can be 
	infinitely close to an equality as $Q\to\infty$.
\end{lemma}
\begin{IEEEproof}
	See Appendix F.
\end{IEEEproof}
%
It is evident from \textbf{Lemmas \ref{lemma:2}-\ref{lemma:rand_MM}} that 
meticulously designed observation matrices can scale down the asymptotic MSE 
from $\mathcal{O}(KMQ^{-1})$ to $\mathcal{O}(K^2Q^{-1})$.
This improvement is attributed to the fact that the observation matrices 
designed by 
water-filling and ice-filling approaches are tightly aligned with the kernel's 
eigenspace 
of dimension $K$. 
In contrast, the random observation matrix wastes a large amount of 
power in the kernel's null space of dimension $M-K$, which contains no 
useful information about the channel. This finding demonstrates the significant 
role played by the observation matrix to enhance channel estimation accuracy, 
particularly in DASs where $M \gg K$.


\subsection{Estimation Accuracy Under Imperfect 
	Kernel}\label{subsec:EA_imperfect_kernel}
We now consider the situation where the perfect prior kernel ${\bm \Sigma}_{\bf 
	h}$ is not available. In this context, an imperfect prior kernel, such as 
	the 
statistical 
kernel in \eqref{eq:statis} and the 
artificial kernels \eqref{eq:exp} and \eqref{eq:bes}, has to be adopted in the 
proposed algorithms. We 
denote the 
adopted prior kernel as $\bf \Sigma$. Then, one can use the 
following lemma to evaluate the MSE performance analytically.
\begin{lemma}\label{lemma:4}
	Given a prior kernel $\bf \Sigma$ as the input of 
	the proposed channel estimator, the MSE $\hat \delta$ of channel estimation 
	is given 
	by
	\begin{align}\label{eqn:nonideal_MSE}
		{\hat \delta}= \mathsf{Tr}\left(\left( {{{\bf{\Pi }}^H}{{\bf{W}}^H} - 
		{{\bf{I}}_M}} \right){{\bf{\Sigma }}_{\bf{h}}}\!\left( {{\bf{W\Pi }} - 
		{{\bf{I}}_M}} \right)\right)  + {\sigma ^2} 	
		\mathsf{Tr}\!\left({{\bf{\Pi }}^H}{\bf{\Pi }} \right),
	\end{align}
	where ${\bf{\Pi }} := {\left( {{{\bf{W}}^H}{\bf{\Sigma W}} + {\sigma 
				^2}{\bf I}_Q} \right)^{ - 1}}{{\bf{W}}^H}{\bf{\Sigma }}$.
\end{lemma}
\begin{IEEEproof}
	See Appendix G.
\end{IEEEproof}
\textbf{Lemma \ref{lemma:4}} indicates that, the estimation error caused by the 
non-ideal input cannot be neglected. For example, if the input kernel $\bm 
\Sigma$ behaves far from the real kernel ${\bm \Sigma}_{\bf h}$, the value of 
the MSE may become significantly large due to the existence of $\bm \Pi$ in 
(\ref{eqn:nonideal_MSE}). 
To obtain more insightful results, we use the 
statistical kernel, ${\bf \hat{\Sigma}}_{\mb{h}}$ in \eqref{eq:statis}, 
as an example to analyze the achievable MSEs.  
To ease the derivation, we first prove a useful expression of the MSE in 
\textbf{Lemma~\ref{lemma:useful_lemma_2}}.
\begin{lemma}\label{lemma:useful_lemma_2}
	The MSE $\hat \delta$ in (\ref{eqn:nonideal_MSE}) under the imperfect 
	kernel 
	${\bf{\hat \Sigma }}_{{\bf h}}$ can be written as a function of 
	${{\bf{W}}{{\bf{W}}^H}}$, given by
	\begin{align}\label{eqn:useful_hat_delta}
		\notag
		\hat \delta  = & \mathsf{Tr}\left( {\left( {{\bf{\hat \Sigma 
		}}_{\bf{h}}^H{{\bf{\Omega }}^H} - {{\bf{I}}_M}} \right){{\bf{\Sigma 
		}}_{\bf{h}}}\left( {{\bf{\Omega }}{{{\bf{\hat \Sigma }}}_{\bf{h}}} - 
		{{\bf{I}}_M}} \right)} \right) \\ &+ {\sigma ^2}\mathsf{Tr}\left( 
		{{{{\bf{\hat \Sigma }}}_{\bf{h}}}{\bf{\Xi }}{{{\bf{\hat \Sigma 
		}}}_{\bf{h}}}} \right),
	\end{align}
	wherein ${\bf{\Omega }}$ and ${\bf{\Xi }}$ are subfunctions of 
	${{\bf{W}}{{\bf{W}}^H}}$, written as
	\begin{align}
		\label{eqn:Omega}
		{\bf{\Omega }} =& {{{\bf{W}}{{\bf{W}}^H}} \over {{\sigma ^2}}} - 
		{{{\bf{W}}{{\bf{W}}^H}} \over {{\sigma ^4}}}{\left( {{\bf{\hat \Sigma 
		}}_{\bf{h}}^{ - 1} + {{{\bf{W}}{{\bf{W}}^H}} \over {{\sigma ^2}}}} 
		\right)^{ - 1}}{\bf{W}}{{\bf{W}}^H}, \\ \notag
		{\bf{\Xi }} =& {{{\bf{W}}{{\bf{W}}^H}} \over {{\sigma ^4}}} - 
		2{{{\bf{W}}{{\bf{W}}^H}} \over {{\sigma ^6}}}{\left( {{\bf{\hat \Sigma 
		}}_{\bf{h}}^{ - 1} + {{{\bf{W}}{{\bf{W}}^H}} \over {{\sigma ^2}}}} 
		\right)^{ - 1}}{\bf{W}}{{\bf{W}}^H} + \\ &{{{\bf{W}}{{\bf{W}}^H}} \over 
		{{\sigma ^8}}} \left({\left( {{\bf{\hat \Sigma }}_{\bf{h}}^{ - 1} + 
		{{{\bf{W}}{{\bf{W}}^H}} \over {{\sigma ^2}}}} \right)^{ - 
		1}}{\bf{W}}{{\bf{W}}^H}\right)^2. 
		\label{eqn:Xi}
	\end{align}
\end{lemma}
\begin{IEEEproof}
	See Appendix H.
\end{IEEEproof}
Recall that the essential difference among the three estimators, i.e., 
water-filling algorithm, ice-filling algorithm, and the estimator with randomly 
generated $\bf W$, is the form of observation matrix $\bf W$. Thus, by 
utilizing {\bf Lemma  \ref{lemma:useful_lemma_2}},  the MSE of different 
algorithms under the imperfect kernel ${\bf{\hat \Sigma }}_{{\bf h}}$ can be 
obtained by replacing ${{\bf{W}}{{\bf{W}}^H}}$ in $\hat \delta$ with their 
corresponding analytical expressions. Aided by some matrix operations, 
we can thereby prove the 
following lemma.
\begin{lemma}\label{lemma:MSE_under_imp_kernel}	
	Under the imperfect kernel ${\bf{\hat \Sigma }}_{{\bf h}}$, the achievable 
	MSE for the water-filling algorithm can be written as
	\begin{align}\label{eqn:MSE_under_imp_kernel_WF}
		\notag
		{{\hat \delta }_{{\rm{wf}}}} &=  {\sigma ^2}\sum\limits_{k = 1}^K 
		{\frac{{{\lambda _k}{\sigma ^2} \!+\! {{\hat p}_k}{{\left( {{\lambda 
		_k} 
								\!+\! \sigma _{\bf{h}}^2} 
								\right)}^2}}}{{{{\left( {{\hat{p}_k}\left( 
								{\lambda_k \!+\! \sigma _{\bf{h}}^2} \right) + 
								{\sigma ^2}} 
							\right)}^2}}}}  \!+\! 
							\sigma^2\!\!\!\!\sum\limits_{m = K + 1}^M 
							{\frac{{{{\hat 
							p}_m}\sigma _{\bf{h}}^4}}{{{{\left( {{{\hat 
							p}_m}\sigma 
								_{\bf{h}}^2 \!+\! {\sigma ^2}} \right)}^2}}}} \\
		&\mathop =\limits^{Q \to  + \infty } {\cal O}\left( {{\sigma 
		^2}{M^2}{Q^{ - 1}}} \right),
	\end{align}
	wherein $\{{\lambda}_1,\cdots,{\lambda}_M\}$ represent the $M$ non-negative 
	eigenvalues of ${\bm \Sigma}_{\bf h}$ in a descending order\footnote{That 
	is to say, if ${\bm \Sigma}_{\bf h}$ is rank-$K$, we have ${\lambda}_m=0$ 
	for all $m>K$.}. The power ${\hat p}_m$ allocated to each eigenvector is 
	calculated by the water-filling principle ${\hat p}_m = \left({\hat \beta} 
	- \frac{\sigma^2}{\lambda_m+\sigma^2_{\bf h}}\right)^{+}$ for 
	$m\in\{1,\cdots,M\}$. The water-level ${\hat \beta}$ can be determined via 
	binary search such that $\sum_{m=1}^{M} {\hat p}_m = Q$. 
	
	Under the imperfect kernel ${\bf{\hat \Sigma }}_{{\bf h}}$, the achievable 
	MSE for the ice-filling algorithm can be written as
	\begin{align}\label{eqn:MSE_under_imp_kernel_IF}
		\notag
		{{\hat \delta }_{{\rm{if}}}} &= {\sigma ^2}\sum\limits_{k = 1}^K 
		{\frac{{{\lambda _k}{\sigma ^2} \!+\! {{\hat n}_k}{{\left( {{\lambda 
									_k} \!+\! \sigma _{\bf{h}}^2} 
									\right)}^2}}}{{{{\left( 
							{{\hat{n}_k}\left( {\lambda_k \!+\! \sigma 
							_{\bf{h}}^2} \right) 
								\!+\! {\sigma ^2}} \right)}^2}}}}  \!+\! 
								\sigma^2\!\!\!\!\sum\limits_{m = K + 1}^M 
		{\frac{{{{\hat n}_m}\sigma _{\bf{h}}^4}}{{{{\left( {{{\hat 
										n}_m}\sigma _{\bf{h}}^2 \!+\! {\sigma 
										^2}} \right)}^2}}}} \\
		&\mathop =\limits^{Q \to  + \infty } {\cal O}\left( {{\sigma 
		^2}{M^2}{Q^{ - 1}}} \right),
	\end{align}
	wherein the pilot reuse frequencies $\{{\hat n}_1,\cdots,{\hat n}_M\}$ can 
	be obtained by {\bf Algorithm \ref{alg:IF}} with ${\bf{\hat \Sigma }}_{{\bf 
	h}}$ being the input kernel. In particular, $\sum_{m=1}^{M} {\hat n}_m = Q$ 
	holds.
	
	For the estimator enabled by a randomly generated $\bf W$ with sufficiently 
	large $Q$, the achievable MSE under the imperfect kernel ${\bf{\hat \Sigma 
	}}_{{\bf h}}$ can be written as
	\begin{align}\label{eqn:MSE_under_imp_kernel_MM}
		\notag
		{{\hat \delta }_{{\rm{rnd}}}} &\approx  {\sigma ^2}\sum\limits_{k = 
		1}^K 
		{\frac{{{\lambda _k}{\sigma ^2} \!+\! \frac{Q}{M}{{\left( {{\lambda _k} 
								\!+\! \sigma _{\bf{h}}^2} 
								\right)}^2}}}{{{{\left( {\frac{Q}{M}\left( 
								{\lambda_k \!+\! \sigma _{\bf{h}}^2} \right) 
								\!+\! {\sigma ^2}} 
							\right)}^2}}}}  \!+\! \sigma^2\!\!\!\! 
							\sum\limits_{m = K + 1}^M 
		{\frac{{\frac{Q}{M}\sigma _{\bf{h}}^4}}{{{{\left( 
							{\frac{Q}{M}\sigma _{\bf{h}}^2 \!+\! {\sigma ^2}} 
							\right)}^2}}}} \\
		&\mathop =\limits^{Q \to  + \infty } {\cal O}\left( {{\sigma 
		^2}{M^2}{Q^{ - 1}}} \right),
	\end{align}
	wherein the approximate equality can be 
	infinitely close to an equality as $Q\to\infty$. 
\end{lemma}
\begin{IEEEproof}
	See Appendix I.
\end{IEEEproof}
From {\bf Lemma \ref{lemma:MSE_under_imp_kernel}}, it is easy to prove that the 
MSEs under the imperfect kernel $\hat{\bf \Sigma}_{\bf h}$ are increased by the 
kernel estimation error $\sigma_{\bf h}^2$. The reason is that, the estimation 
error makes the input kernel full-rank, i.e., ${\hat {\bf{\Sigma 
	}}_{\bf{h}}} = 
{{\bf{\Sigma }}_{\bf{h}}} + \sigma _{\bf{h}}^2{{\bf{I}}_M}$. Then, the 
power (or pilot reuse frequency) for estimation may be allocated to all 
eigenvectors of ${\bf \Sigma}_{\bf 
	h}$. However, allocating power to those eigenvectors associated with the 
eigenvalue $\sigma _{\bf{h}}^2$ contributes nearly no improvement to channel 
recovery 
accuracy. Besides, an 
imperfect kernel input also changes the weights of the posterior mean ${\bm 
	\mu}_{{\bf h}|{\bf y}}$ in (\ref{eqn:bayesian}), which resembles a Bayesian 
estimator that underestimates the impact of noise. 

\begin{remark}
	\emph{
		According to (\ref{eqn:MSE_under_imp_kernel_WF}), 
		(\ref{eqn:MSE_under_imp_kernel_IF}), and 
		(\ref{eqn:MSE_under_imp_kernel_MM}), an efficient way to alleviate the 
		effect of $\sigma^2_{\bf h}$ on MSEs is to introduce the prior 
		knowledge that the real kernel $\bf{\Sigma}_{\bf h}$ is rank-$K$. Then, 
		the estimator only allocates power to the eigenvalues associated with 
		the $K$-largest eigenvalues of ${\hat {\bf{\Sigma }}_{\bf{h}}}$, so 
		that the second term in (\ref{eqn:MSE_under_imp_kernel_WF}), 
		(\ref{eqn:MSE_under_imp_kernel_IF}), and 
		(\ref{eqn:MSE_under_imp_kernel_MM}) can be eliminated. For example, 
		when $p_m$ for $m>K$ are forced to be zero, ${{\hat \delta 
			}_{{\rm{wf}}}}$ in (\ref{eqn:MSE_under_imp_kernel_WF}) becomes
		\begin{align}\label{eqn:MSE_under_imp_kernel_WF_force}
			\notag
			{\hat \delta }_{{\rm{wf}}} &= {\sigma ^2}\sum\limits_{k = 1}^K 
			{{{{\lambda _k}{\sigma ^2} + {p_k}{{\left( {{\lambda _k} + \sigma 
									_{\bf{h}}^2} \right)}^2}} \over {{{\left( 
									{{p_k}\left( {{\lambda 
											_k} + \sigma _{\bf{h}}^2} \right) + 
											{\sigma ^2}} \right)}^2}}}} \\ 
			&\mathop = \limits^{Q \to  + \infty } {\cal O}\left( {{\sigma 
					^2}{K^2}{Q^{ - 1}}} \right), 
		\end{align} 
		wherein $p_k$ for $k\in\{1,\cdots,K\}$ is given in (\ref{eq:WF}). For 
		$\sigma _{\bf{h}}^2>0$, it is evident that ${\hat \delta 
		}_{{\rm{wf}}}$ in
		(\ref{eqn:MSE_under_imp_kernel_WF_force}) is lower than that in 
		(\ref{eqn:MSE_under_imp_kernel_WF}).
	}
\end{remark}

Based on the results in {\bf Lemma \ref{lemma:MSE_under_imp_kernel}}, the 
asymptotic MSEs when the kernel estimation error 
$\sigma_{\mb{h}}^2$ is infinitely large are expressed as follows. 

\begin{corollary}\label{corollary:5}
	When $\sigma _{\bf{h}}^2 \to  + \infty $, the asymptotic MSEs ${\hat 
		\delta}$ for the water-filling algorithm, ice-filling algorithm, and 
		the 
	estimator based on random $\bf W$ can be respectively written as
	\begin{align}
		{\hat \delta}_{\rm wf} &\mathop  = \limits^{\sigma _{\bf{h}}^2 \to  + 
			\infty }  {\cal O}\left( {{\sigma ^2}\sum\nolimits_{k = 
				1}^{{L_{{\rm{wf}}}}} {\hat p_k^{ - 1}} } \right) = {\cal 
				O}\left( 
		{{\sigma ^2}{M^2}{Q^{ - 1}}} \right), \\
		{\hat \delta}_{\rm if} &\mathop  = \limits^{\sigma _{\bf{h}}^2 \to  + 
			\infty }  {\cal O}\left( {{\sigma ^2}\sum\nolimits_{k = 
				1}^{{L_{{\rm{if}}}}} {\hat n_k^{ - 1}} } \right) = {\cal 
				O}\left( 
		{{\sigma ^2}{M^2}{Q^{ - 1}}} \right), \\
		{\hat \delta}_{\rm rnd}  &\mathop  = \limits^{\sigma _{\bf{h}}^2 \to  + 
			\infty } {\cal O}\left( {{\sigma ^2}{M^2}{Q^{ - 1}}} \right),
	\end{align}
	where $L_{\rm wf}$ and $L_{\rm if}$ are the numbers of the positive values 
	in $\{{\hat p}_m\}_{m=1}^{M}$ and $\{{\hat n}_m\}_{m=1}^{M}$, respectively.
\end{corollary}
\begin{IEEEproof}
	The three asymptotic MSEs can be easily obtained by letting $\sigma 
	_{\bf{h}}^2 \to  + \infty $ for ${\hat \delta}_{\rm wf}$ in 
	(\ref{eqn:MSE_under_imp_kernel_WF}), ${\hat \delta}_{\rm if}$ in 
	(\ref{eqn:MSE_under_imp_kernel_IF}), and ${\hat \delta}_{\rm rnd}$ in 
	(\ref{eqn:MSE_under_imp_kernel_MM}), respectively. Since their derivations 
	are similar, here we focus on the asymptotic ${\hat \delta}_{\rm wf}$ as an 
	example. By letting $\sigma _{\bf{h}}^2 \to  + \infty $ and removing the 
	small-order components, we have ${\hat \delta}_{\rm wf} \mathop  = 
	\limits^{\sigma _{\bf{h}}^2 \to  + \infty }  {\cal O}\left( {{\sigma 
			^2}\sum\nolimits_{k = 1}^{{L_{{\rm{wf}}}}} {\hat p_k^{ - 1}} } 
			\right)$. 
	Since ${\hat p}_m = \left({\hat \beta} - 
	\frac{\sigma^2}{\lambda_m+\sigma^2_{\bf h}}\right)^{+}$ for 
	$m\in\{1,\cdots,M\}$, we have ${\hat p}_m = {\hat \beta} = \frac{Q}{M}$ and 
	$L_{{\rm{wf}}} = M$, which completes the proof.
\end{IEEEproof}
\subsection{Computational Complexity Analysis}
The computational complexity of the proposed channel estimators is low in 
practical applications. Specifically, according to Fig. \ref{img:framework}, 
the computational complexity can be divided into two components, i.e., 
observation matrix design in Stage 1 and Bayesian regression in Stage 2. 

For the observation matrix design, the complexity of \textbf{Algorithm 
	\ref{alg:IF}} is 
mainly 
caused by the eigenvalue decomposition and the principal eigenvalue selection, 
which is 
${\cal O}(M^3+QK)$. The complexity of \textbf{Algorithm \ref{alg:mm}} mainly 
depends on 
the search for the largest eigenvalue and the iterations required for the 
convergence of MM, which is ${\cal O}(M^3+I_oM^2Q)$ {\color{black} wherein 
$I_o$ 
	is the 
	number of iterations.} 

For the Bayesian regression, according to (\ref{eqn:bayesian}), we find that 
the recovered channel $\hat{\bf h}$ is exactly the weighted sum of received 
pilots $\bf y$. In particular, the weight is calculated by ${\bf{\Sigma 
	}_{\bf{h}}}{{\bf{W}}}{\left( {\bf W}^{H}{\bf{\Sigma }}_{\bf{h}}{\bf W} + 
	\sigma^2 {\bf I}_Q \right)^{-1}}$, thus the overall complexity of Stage 2 
	is 
${\cal O}(M^3)$. It is notable that the weight only relies on the designed 
${\bf W}$ and $\bm{\Sigma}_{\bf h}$. It means that the weight for recovering 
$\bf h$ can also be calculated offline and then employed for online channel 
estimation.

\section{Simulation Results}\label{sec:6}
\begin{table}[t]\color{black}
	\centering
	\small
	\caption{Simulation Parameters of Channel Model in 3GPP TR 38.901}
	\label{table:2}
	\setstretch{1.25}
	\begin{tabular}{|c|c|c|c|c|c|c|c|}
		\hline  
		\bf Channel parameters&\bf Values in \cite{cdl}\\
		\hline  
		Carrier frequency $f_c$ & 3.5 GHz \\ \hline
		Number of clusters & 23 \\ \hline
		Number of rays per cluster  & 20 \\ \hline
		Path gains & ${\cal CN}(0,1)$  \\ \hline
		Incident angles & ${\cal U}(-90^\circ,+90^\circ)$ \\ \hline
		Maximum angle spread & ${\cal U}(-5^\circ,+5^\circ)$ \\ \hline
		Maximum delay spread & ${\cal U}(-30~{\rm ns},+30~{\rm ns})$ \\ \hline
	\end{tabular}
	\vspace*{-1em}
\end{table}
In this section, we present simulation results to verify the effectiveness of 
the proposed channel estimators for \acp{das}. 
\subsection{Simulation Setup and Benchmarks}
To account for a practical scenario, the 3GPP TR 38.901 channel model 
is used for simulations, whose key parameters are given in Table \ref{table:2}. 
Otherwise specifically specified, we consider a UPA. The 
number of antennas is set to $M=128$, and the numbers of 
horizontal antennas 
and vertical antennas are set to $M_x = 16$ and $M_y = 8$, respectively. The 
antenna spacing is set to $\frac{\lambda}{8}$.  The \ac{snr} is defined as 
${\rm 
	SNR}=\frac{{\mathsf E}\left(\|{\bf h}\|^2\right)}{\sigma^2}$. 
The performance is evaluated by the \ac{nmse}, which is 
defined as ${\rm NMSE}={\mathsf E}\left(\frac{\|{\bf h}-\hat{\bf h}\|^2}{\|{\bf 
		h}\|^2}\right)$. 3000 Monte-Carlo experiments are carried out to plot 
each figure.

{\color{black} 
	To demonstrate the effectiveness of the proposed ice-filling and MM 
	algorithms, 
	six observation matrices are considered for comparison:
	\begin{itemize}
		\item Random matrix $\mb{W}^{\rm Rand} \in\mathbb{C}^{M\times Q}$: The 
		coefficients of $\mb{W}^{\rm Rand}$ are independently generated from  
		the
		standard complex Gaussian 
		distribution. Then, each column is normalized to satisfy the unit power 
		constraint. 
		\item Top-$Q$ matrix $\mb{W}^{{\rm Top}Q} \in\mathbb{C}^{M\times Q}$: 
		The 
		first $Q$ major eigenvectors of ${\bf \Sigma}_{\mb{h}}$ are used to set 
		the 
		observation matrix 
		$\mb{W}^{{\rm Top}Q}$. 
		\item Ice-filling matrix $\mb{W}^{\rm IF} \in\mathbb{C}^{M\times Q}$: 
		The 
		proposed ice-filling algorithm is used to generate  $\mb{W}^{\rm IF}$. 
		\item MM matrix $\mb{W}^{\rm MM} \in\mathbb{C}^{M\times Q}$: The 
		proposed 
		MM algorithm is used to generate $\mb{W}^{\rm MM}$.
		\item Water-filling matrix $\mb{W}^{\rm WF} \in\mathbb{C}^{M\times Q}$: 
		The 
		water-filling scheme is employed to 
		generate  $\mb{W}^{\rm WF}$. To implement 
		this ideal observation matrix, the noise 
		vector, $\mb{z}$, 
		in 
		\eqref{eqn:y} is assumed to be independent of the observation 
		matrix $\mb{W}^{\rm WF}$, and its distribution is fixed as 
		$\mathcal{CN}(\mb{0}_Q, 
		\sigma^2 \mb{I}_Q)$. 
		\item DFT matrix $\mb{W}^{\rm DFT}\in\mathbb{C}^{M\times M}$: The DFT 
		matrix requires the pilot length, $Q$, to be equal to the number of 
		antennas, $M$. This matrix is used for performing LS algorithm.
	\end{itemize}
	
	Note that the pilot length for the DFT matrix $\mb{W}^{\rm DFT}$ is $Q = M 
	= 
	128$, while that for the other matrices is $Q = 64$ by default. The above 
	observation matrices are used in four channel estimators for comparison: 
	The 
	LS, vector approximate message passing (VAMP)~\cite{rangan2019vector}, OMP, 
	and 
	MMSE algorithms.  
	For the LS algorithm, it has to be implemented by the DFT matrix 
	$\mb{W}^{\rm 
		DFT}$ with a $Q = M$ pilot length. As for the OMP and VAMP algorithms, 
		we 
	conduct numerical experiments in advance to evaluate their performance under
	matrices, $\mb{W}^{\rm Rand}$, $\mb{W}^{{\rm Top}Q}$, $\mb{W}^{\rm MM}$ and 
	$\mb{W}^{\rm IF}$, 
	respectively. It is shown that the combinations of ``VAMP + $\mb{W}^{\rm 
		Rand}$" and ``OMP + $\mb{W}^{{\rm Top}Q}$" exhibit the best channel 
	estimation performance\footnote{This is because CS algorithms expect to 
	make 
		the column coherence of observation matrix as small as possible. As the 
		designed matrix $\mb{W}^{{\rm IF}}$ might select the same eigenvectors 
		in 
		different time slots, it is 
		not suitable for OMP and VAMP algorithms.}.  Thereby, the following 
		simulations 
	use the random matrix $\mb{W}^{\rm Rand}$ for VAMP and the top-$Q$ matrix 
	$\mb{W}^{{\rm Top}Q}$ for 
	OMP. The MMSE algorithm in \eqref{eqn:bayesian} is the estimator adopted in 
	this paper. We will evaluate its performance under matrices, $\mb{W}^{\rm 
		Rand}$,  $\mb{W}^{{\rm Top}Q}$,  $\mb{W}^{\rm MM}$, $\mb{W}^{\rm IF}$, 
		and  
	$\mb{W}^{\rm WF}$, respectively, to show the superiority of the designed 
	observation matrices.  
}


\subsection{Performance Evaluation Under Perfect Kernel}
\begin{figure}[!t]
	\centering
	\includegraphics[width=0.47\textwidth]{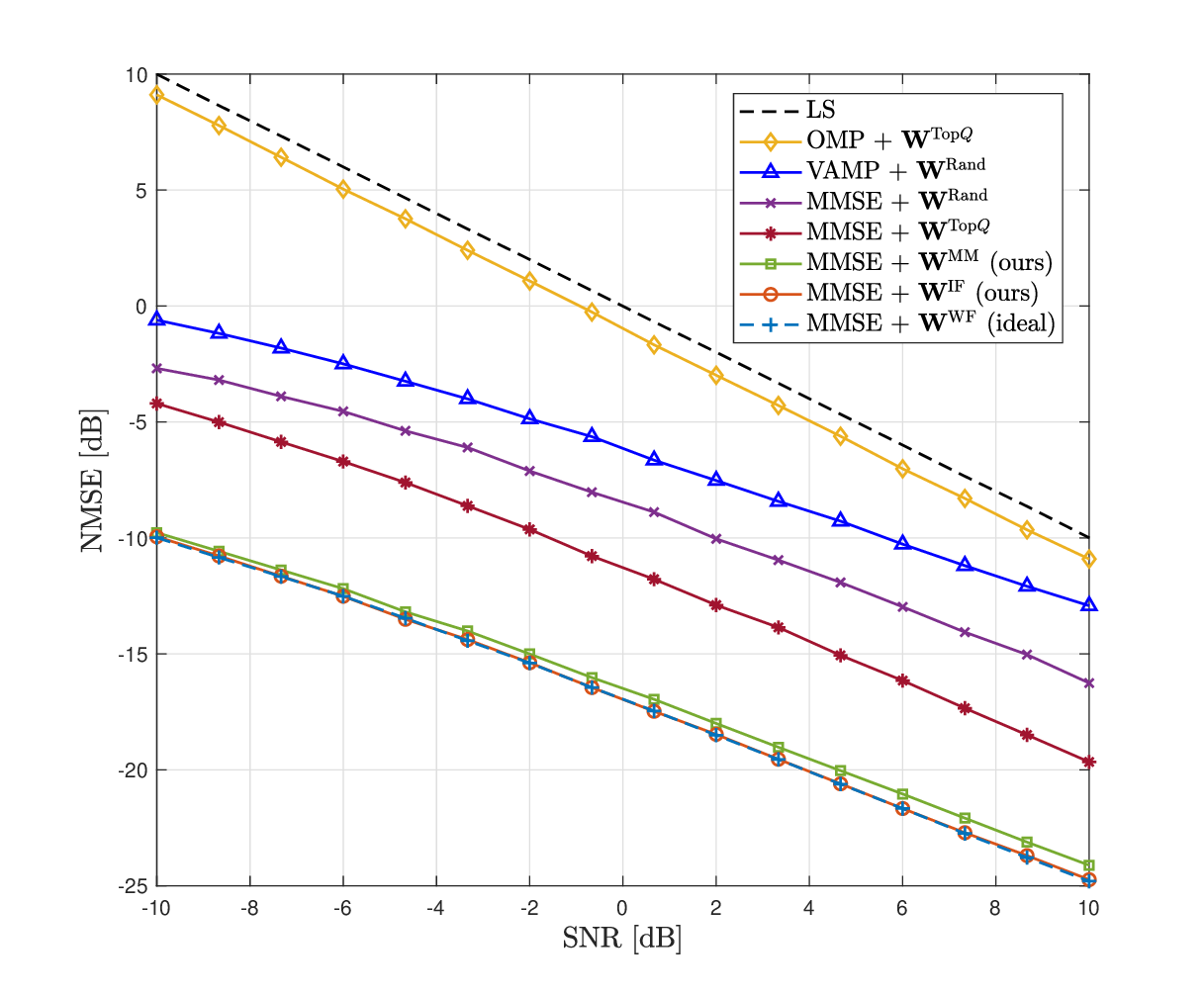}
	\vspace*{-0.5em}
	\caption{The effect of SNR on NMSE performance under perfect kernel
		${\bm \Sigma}_{\bf h}$. 
	}
	\vspace*{-1em}
	\label{img:NMSE_SNR_PK}
\end{figure}

In this subsection, we use the perfect kernel, ${\bm \Sigma}_{\mb{h}}$, to 
trigger the water-filling, ice-filling, and MM algorithms, and then perform the 
Bayesian regression. 
The NMSE performance 
as a 
function of SNR  is evaluated in Fig.~\ref{img:NMSE_SNR_PK}. The SNR ranges 
from 
$-10{\:}{\text{dB}}\sim10{\:}{\text{dB}}$. Thanks to the careful design of 
observation matrix, our schemes ``MMSE + $\mb{W}^{\rm MM}$" and ``MMSE + 
$\mb{W}^{\rm IF}$" remarkably outperform existing benchmarks. For example, a 
10~dB NMSE gap between ``MMSE + $\mb{W}^{\rm IF}$" and 
``VAMP + $\mb{W}^{\rm Rand}$ is visible. Moreover, the ice-filling matrix  
$\mb{W}^{\rm IF}$ can reduce the NMSE of MMSE algorithms by 5 dB compared to 
the 
top-$Q$ matrix $\mb{W}^{{\rm Top}Q}$. Notably, the NMSE performance of MMSE 
attained by 
the ice-filling matrix 
tightly aligns with that of the ideal water-filling matrix, given that the 
ice-filling 
intrinsically allocates $Q$ quantized units of power  to the $Q$-timeslot 
pilots. 
Another 
finding of interest is that the NMSE gap between ``MMSE + $\mb{W}^{\rm IF}$" 
and ``MMSE + $\mb{W}^{\rm MM}$" is 
no higher than 0.5 dB, demonstrating the near-optimality of the proposed MM 
algorithm in phase-only-controllable design. 

\begin{figure}[!t]
	\centering
	\includegraphics[width=0.47\textwidth]{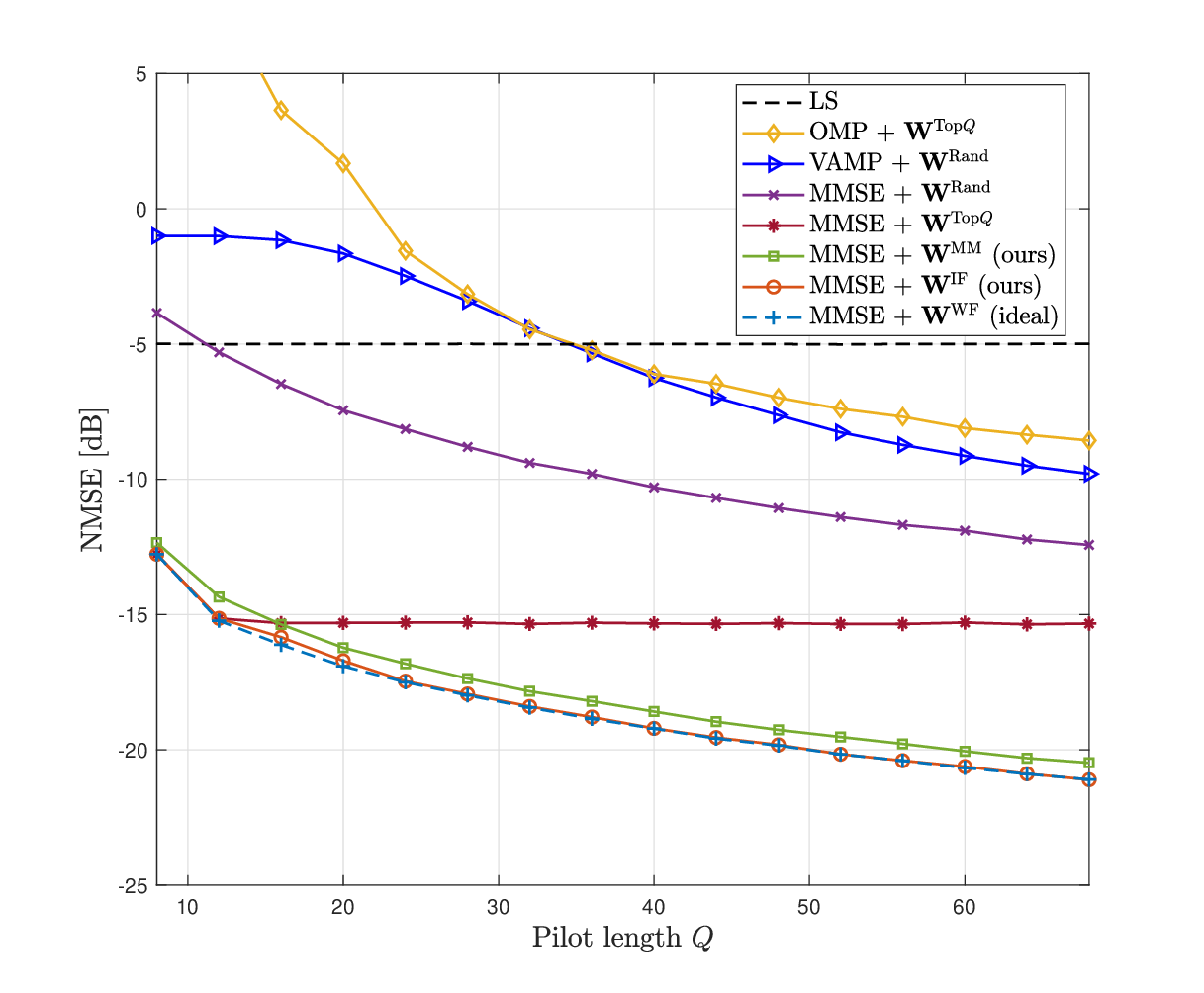}
	\vspace*{-0.5em}
	\caption{The NMSE as a function of pilot length. The pilot length allocated 
		to the LS method is fixed to 128, whereas the pilot length $Q$ for 
		other schemes rises from 8 to 68.
	}
	\vspace*{-1em}
	\label{img:NMSE_Pilot_PK}
\end{figure}
\begin{figure}[!t]
	\centering
	\includegraphics[width=0.47\textwidth]{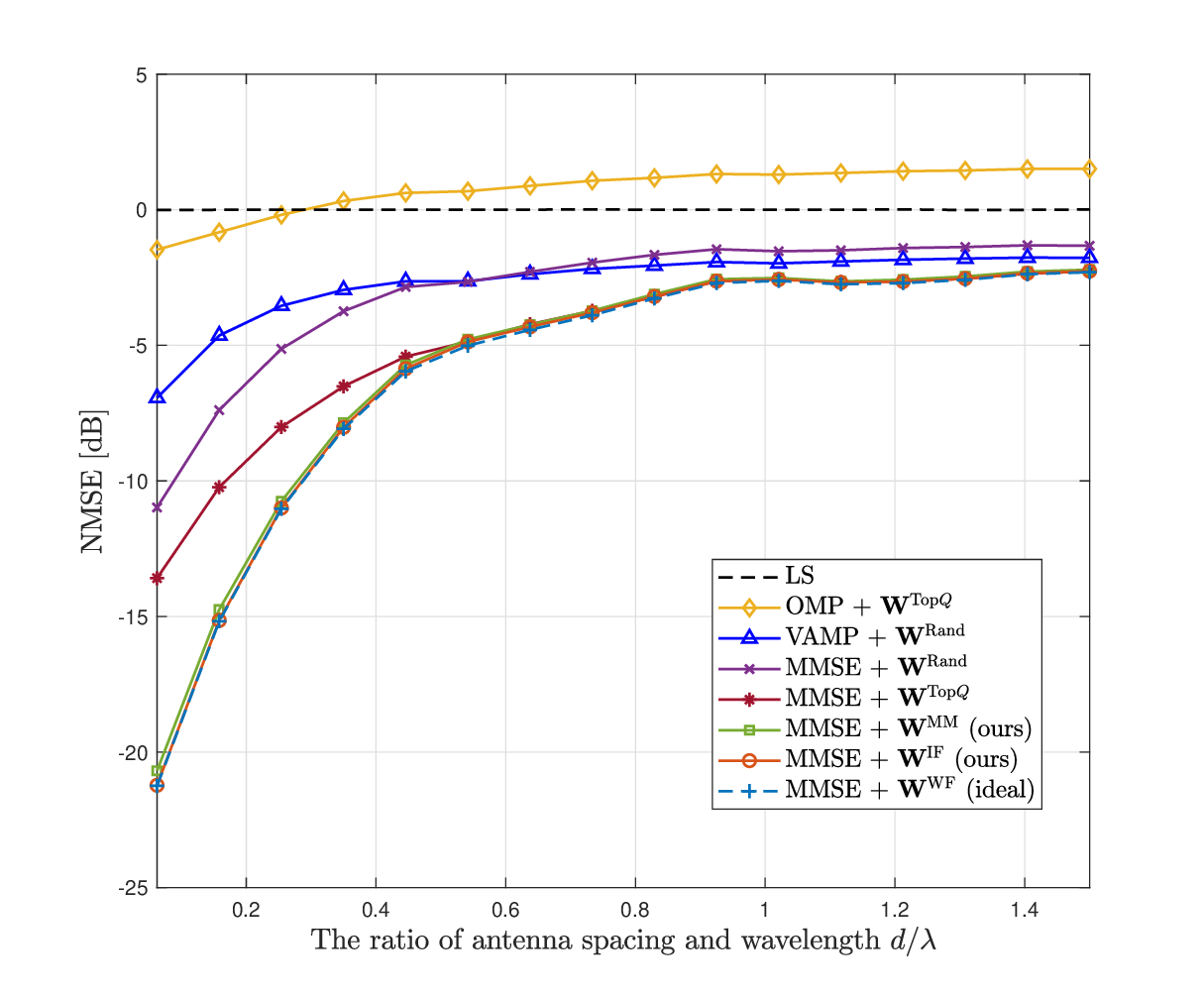}
	\vspace*{-0.5em}
	\caption{NMSE versus the ratio of antenna spacing and wavelength. 
	}
	\vspace*{-1em}
	\label{img:NMSE_Spacing_PK}
\end{figure}
Next, we investigate the impact of pilot length, $Q$, on the NMSE performance 
in 
Fig.~\ref{img:NMSE_Pilot_PK}. The pilot length allocated to the LS 
method is fixed to 128, whereas the pilot length for the other algorithms 
ranges 
from 
8 to 68. The SNR is -5 dB. It is evident from Fig.~\ref{img:NMSE_Pilot_PK} that 
the 
proposed 
algorithms 
consume significantly lower pilot overhead than the other benchmarks to achieve 
the same 
NMSE level. For instance, to obtain a $-5\,\text{dB}$ NMSE, the pilot length 
required  
for the VAMP algorithm is 
larger than 30. In contrast,  8 pilots are sufficient for the 
proposed ice-filling and MM algorithms to attain a NMSE lower than - 10 dB. 
Additionally, it is notable that the NMSE performance of ``MMSE + $\mb{W}^{{\rm 
		Top}Q}$ no longer decreases when $Q > 12$. This is because 
the top-$Q$ matrix begins to allocate pilots to invalid eigenvectors with very 
small eigenvalues when $Q > 12$, while the proposed ice-filling and MM 
algorithms continue to select valid eigenvectors from the $1^{\rm st}$ to 
$12^{\rm th}$ eigenvalues. As a result, the performance gain attained by our 
algorithms continuously improve as the pilot length increases. 



In Fig.~\ref{img:NMSE_Spacing_PK}, we demonstrate the superiority of a DAS in 
precise channel estimation. The curves of the achieved NMSE 
performance versus the ratio of antenna spacing and wavelength, i.e., $
\frac{d}{\lambda}$, are 
plotted. The SNR is 0 dB. We can observe that, as the normalized antenna 
spacing 
$\frac{d}{\lambda}$ decreases from $\frac{3}{2}$ to $\frac{1}{16}$, the NMSEs 
for all 
channel estimators excluding LS method declines rapidly.  For example, the NMSE 
achieved by the proposed ice-filling algorithm is 
decreased by about 15 dB when the antenna spacing ranges from $\lambda/2$ to 
$\lambda/8$.
This phenomenon is attributed to the fact that, a smaller antenna spacing leads 
to stronger spatial correlations of channels over different antennas, which 
results in a more informative kernel ${\bm \Sigma}_{\mb{h}}$ for more precise 
channel reconstruction. Since the underpinning spatial 
information of wireless channels is embedded in the structured kernel, these  
Bayesian regression based channel estimators are enhanced under DASs. 

\subsection{Performance Evaluation Under Imperfect Kernel}

\begin{figure}[!t]
	\centering
	\includegraphics[width=0.47\textwidth]{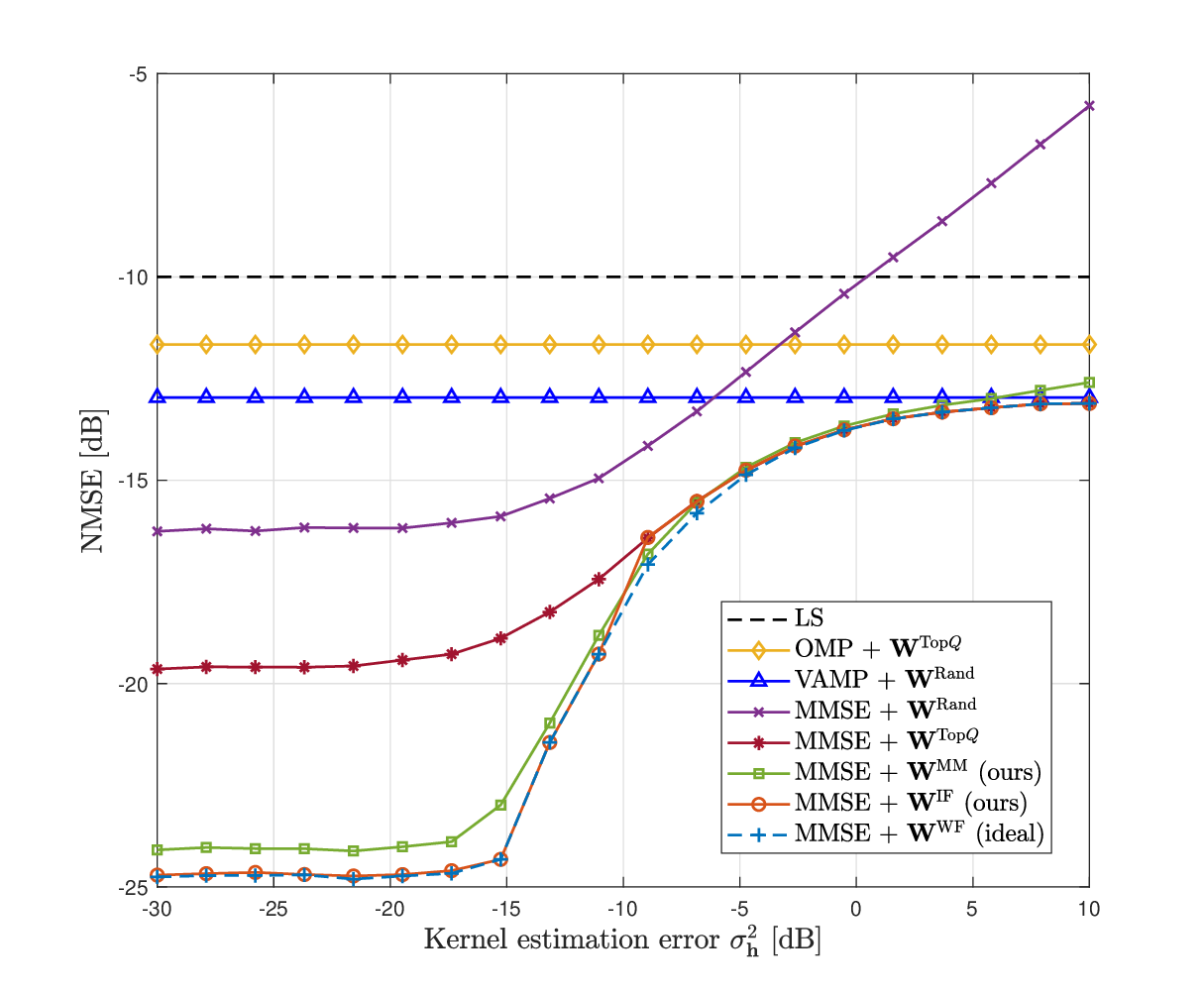}
	\vspace*{-0.5em}
	\caption{The effect of kernel estimation error on NMSE performance under 
		the imperfect 
		kernel 	$\hat{\bm \Sigma}_{\bf h}$. 
	}
	\vspace*{-1em}
	\label{img:NMSE_CE_IPK}
\end{figure}
In some scenarios where the perfect prior kernel $\bm{\Sigma}_{\bf h}$ cannot 
be obtained, 
the statistical kernel and 
artificial kernels defined in Section \ref{subsec:4:D} can be utilized to 
replace the perfect kernel 
$\bm{\Sigma}_{\bf h}$ in Algorithms \ref{alg:IF}$\sim$\ref{alg:mm}. We first  
consider the statistical kernel ${\bf{\hat \Sigma }}_{{\bf	h}} = {\bf{ \Sigma 
}}_{{\bf h}} + 
\sigma^2_{\mb{h}}\mb{I}_M$. The NMSE as a function of the kernel estimation 
error $\sigma^2_{\mb{h}}$ is shown in Fig. 
\ref{img:NMSE_CE_IPK}. One can observe that all MMSE methods suffer from 
visible performance loss due to the imperfect kernel input. Fortunately, the 
proposed methods still hold the superiority among all schemes. In particular, 
the estimation accuracy of the proposed methods is almost not influenced when 
$\sigma^2_{\mb{h}}<-15\,{\text{dB}}$, which shows their high robustness to the 
kernel error. 

\begin{figure}[!t]
	\centering
	\includegraphics[width=0.47\textwidth]{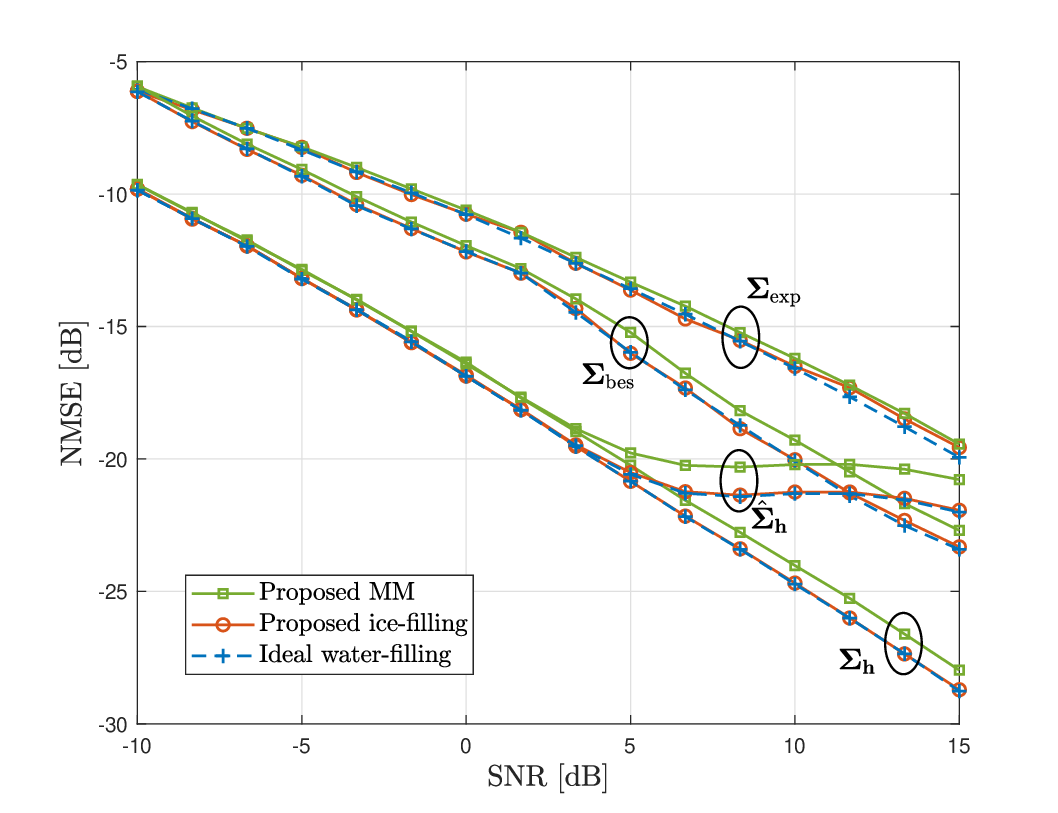}
	\vspace*{-0.5em}
	\caption{The effect of SNR on NMSE performance under the perfect kernel 
		${\bm \Sigma}_{\bf h}$, the statistical 
		kernel $\hat{{\bm \Sigma}}_{\mb{h}}$, the Bessel kernel ${\bm 
			\Sigma}_{\rm bes}$, and 
		the 
		exponential kernel ${\bm \Sigma}_{\rm exp}$.
	}
	\vspace*{-1em}
	\label{img:NMSE_SNR_IPK}
\end{figure}

\begin{figure}[!t]
	\centering
	\includegraphics[width=0.47\textwidth]{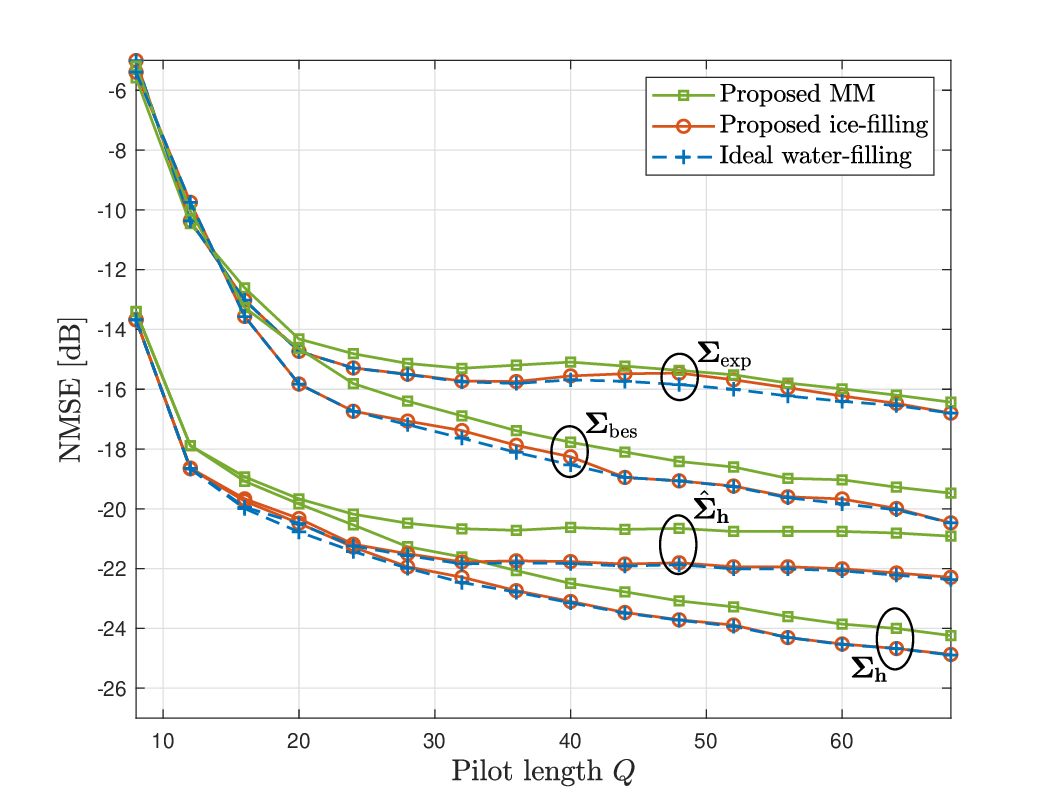}
	\vspace*{-0.5em}
	\caption{The effect of pilot length on NMSE performance under the perfect 
		kernel 
		${\bm \Sigma}_{\bf h}$, the Bessel kernel ${\bm \Sigma}_{\rm bes}$, and 
		the 
		exponential kernel ${\bm \Sigma}_{\rm exp}$. 
	}
	\vspace*{-1em}
	\label{img:NMSE_Pilot_IPK}
\end{figure}

We further evaluate the achievable NMSE performance under all considered 
kernels, including the perfect kernel ${\bm \Sigma}_{\mb{h}}$, the statistical 
kernel $\hat{{\bm \Sigma}}_{\mb{h}}$, and the artificial kernels 
$\bm{\Sigma}_{\rm exp}$ and ${\bm \Sigma}_{\rm bes}$.  The estimation error of 
$\hat{{\bm \Sigma}}_{\mb{h}}$ is set as $\mb{\sigma}_{\mb{h}}^2 = - 12\:{\rm 
	dB}$. 
The hyper-parameters, $\eta_1$ and $\eta_2$, of $\bm{\Sigma}_{\rm exp}$ and 
${\bm \Sigma}_{\rm bes}$ are determined by the binary 
search 
through several 
numerical experiments, which are set to $\eta_1 = 0.56$ and $\eta_2=0.85$ in 
the considered scenario. 
In this way, we obtain the NMSE versus the SNR and that versus the pilot length 
$Q$ in Fig. \ref{img:NMSE_SNR_IPK} and Fig. \ref{img:NMSE_Pilot_IPK}, 
respectively. For clarity, we only plot the curves of MMSE algorithms. 
From these two figures one find that, for a given kernel, the curves of 
``ideal water-filling'', ``proposed ice-filling'', and ``proposed MM'' are very 
close, which indicates that 
these 
three schemes have similar robustness to imperfect kernels. 

More importantly, one can find that the imperfect kernels lead to a decrease in 
estimation 
performances, while the CSI accuracy still holds considerable. Compared with 
the NMSE achieved by the perfect kernel, those achieved by the imperfect 
kernels experience 
an increase of about $5{\:}\text{dB}$. For example, when the SNR is set to 
$5{\:}\text{dB}$, the NMSEs of the proposed estimators for kernels 
$\bm{\Sigma}_{\bf h}$, $\hat{\bm{\Sigma}}_{\bf h}$, ${\bm \Sigma}_{\rm bes}$, 
and $\bm{\Sigma}_{\rm exp}$ 
are about $-20{\:}\text{dB}$, $-20{\:}\text{dB}$, $-15{\:}\text{dB}$, and 
$-13{\:}\text{dB}$, 
respectively. To achieve the NMSE of $-15{\:}\text{dB}$, the required pilot 
lengths $Q$ for the four kernels are $10$, $10$, $18$, and $20$, respectively. 
Despite the 
performance losses, we find that the superiority of the proposed methods still 
hold while comparing to the baselines in Fig. \ref{img:NMSE_SNR_PK} and Fig. 
\ref{img:NMSE_Pilot_PK}. This observation is encouraging because it  suggests 
that the proposed methods may not require real prior knowledge of channels. 
As an alternative, a virtual kernel composed of experiential parameters may be 
an ideal choice in practice, such that gains without pain can be achieved. 

\section{Conclusions}\label{sec:7}
This paper incorporated the design of observation matrix into Bayesian channel 
estimation in DASs.  
The formulated MIM-based observation matrix design was shown to be a 
time-domain duality of classic MIMO precoding, which was ideally addressed by 
the water-filling principle. Targeting practical DAS realizations, we proposed 
a novel ice-filling and a MM enabled algorithms to design amplitude-and-phase 
controllable and phase-only controllable observation matrices, respectively. In 
particular, 
the ice-filling algorithm was proved to be a discrete approximation of 
water-filling, 
with the relative quantization error decaying rapidly. Comprehensive 
analyses and numerical results  
validated the 
near-optimality of the proposed designs. 

This work establishes a new understanding of channel estimation from the view 
of information theory. Several potential directions for extending our work are 
summarized as follows. In current communication systems, beam selection 
techniques, including beam training and beam tracking, are widely adopted CSI 
acquisition approaches. Designing new beam selection strategies using the idea 
of MIM will be interesting. 
Besides, the extension to more general systems, such as multi-user MIMO and 
wideband communications,  and theoretically analyzing their channel estimation 
accuracy also deserve in-depth study. 




\begin{appendices}
	\section{Proof of \eqref{eq:lemma1}}\label{appendix:1}
	The posterior kernel $\boldsymbol{\Sigma}_t$ is expressed as 
	\begin{align}\label{eq:kernel}
		{{\bf{\Sigma }}_t} = {{\bf{\Sigma }}_{\bf{h}}} - {{\bf{\Sigma 
			}}_{\bf{h}}}{{\bf{W}}_t}{
			\left( {{{\bf{W}}_t^H} {{{\bf{\Sigma 
						}}_{\bf{h}}}{{ {{{\bf{W}}_t}} }} + {\sigma 
						^2}{{\bf{I}}_{t}}} } \right)^{ - 
				1}}
		{ {{{\bf{W}}_t^H}} }
		{{\bf{\Sigma }}_{\bf{h}}}. 
	\end{align}
	The key to the proof of \eqref{eq:lemma1} lies in calculating the inverse 
	of ${{{\bf{W}}_t^H} 
		{{{\bf{\Sigma 
				}}_{\bf{h}}}{{ {{{\bf{W}}_t}} }} + {\sigma 
				^2}{{\bf{I}}_t}} }$ in (\ref{eq:kernel}). For ease of 
	notation, we denote ${\bf{K}}_t = {{{\bf{W}}_t^H} {{{\bf{\Sigma 
				}}_{\bf{h}}}{{ {{{\bf{W}}_t}} }} + {\sigma 
				^2}{{\bf{I}}_t}} }$ and ${\bf{W}}_t = 
	[{{\bf{W}}_{t-1}}, {\bf{w}}_t]$. Then ${\bf{K}}_t$ 
	can be expressed in a block matrix form:
	\begin{align} \label{eq:L1p1}
		{\bf{K}}_t =& \left[
		\begin{array}{cc}
			{{{\bf{W}}_{t-1}^H} {{{\bf{\Sigma 
						}}_{\bf{h}}} {{ {{{\bf{W}}_{t-1}}} }} + {\sigma 
						^2}{{\bf{I}}_{t-1}}} }	&  
			{{\bf{W}}_{t-1}^H}{{\bf{\Sigma 
				}}_{\bf{h}}} {\bf{w}}_t
			\\ 
			{\bf{w}}_t^H{{\bf{\Sigma 
				}}_{\bf{h}}} {{\bf{W}}_{t-1}}	&	
			{\bf{w}}_t^H{{\bf{\Sigma 
				}}_{\bf{h}}} {\bf{w}}_t + \sigma^2
		\end{array}\right] \notag \\ = &
		\left[
		\begin{array}{cc}
			{\bf{K}}_{t-1}	&  {{\bf{W}}_{t-1}^H}{{\bf{\Sigma 
				}}_{\bf{h}}} {\bf{w}}_t
			\\ 
			{\bf{w}}_t^H{{\bf{\Sigma 
				}}_{\bf{h}}} {{\bf{W}}_{t-1}}	&	
			{\bf{w}}_t^H{{\bf{\Sigma 
				}}_{\bf{h}}} {\bf{w}}_t + \sigma^2
		\end{array}\right].
	\end{align}
	We can harness the block-matrix inversion in (\ref{eq:L1p2})
	\begin{figure*}[!b]
		\normalsize	
		\hrulefill
		\begin{align} \label{eq:L1p2}
			\left[
			\begin{array}{cc}
				{\bf A}	&  	{\bf B}
				\\ 
				{\bf C}	&		{\bf D}
			\end{array}\right]^{-1} = 	\left[
			\begin{array}{cc}
				{\bf A}^{-1} + {\bf A}^{-1} {\bf B}({\bf D} - {\bf C }{\bf 
					A}^{-1} {\bf 
					B})^{-1}{\bf C}{\bf A}^{-1}	&  	-{\bf A}^{-1}{\bf B}({\bf 
					D} - {\bf C}{\bf 
					A}^{-1}{\bf B})^{-1}
				\\ 
				-({\bf D} - {\bf C}{\bf A}^{-1}{\bf B})^{-1}{\bf C} {\bf 
					A}^{-1}	
				&			({\bf D} - {\bf C}{\bf A}^{-1}{\bf 
					B})^{-1} 
			\end{array}\right]
		\end{align} 
	\end{figure*}
	to derive the inverse of ${\bf{K}}_t$~\cite{Meyer2000Matrix}. By mapping 
	the 
	matrices $\mb{A}$, 
	$\mb{B}$, 
	$\mb{C}$, and $\mb{D}$ in \eqref{eq:L1p1} to \eqref{eq:L1p2}, the value 
	of 
	${\bf 
		D} - {\bf C }{\bf A}^{-1} 
	{\bf 
		B}$ is expressed as
	\begin{align} \label{eq:L1p3}
		&{\bf 
			D} - {\bf C }{\bf A}^{-1} 
		{\bf 
			B} \notag \\ &= {\bf{w}}_t^H{{\bf{\Sigma 
			}}_{\bf{h}}} {\bf{w}}_t + \sigma^2 - {\bf{w}}_t^H{{\bf{\Sigma 
			}}_{\bf{h}}} {{\bf{W}}_{t-1}} {\bf{K}}_{t-1}^{-1} 
		{{\bf{W}}_{t-1}^H}{{\bf{\Sigma 
			}}_{\bf{h}}} {\bf{w}}_t \notag \\
		& \overset{(a)}{=} {\bf{w}}_t^H {\bf \Sigma}_{t-1} {\bf{w}}_t + 
		\sigma^2,
	\end{align}
	where $(a)$ comes from the definition of ${\bf \Sigma}_{t-1}$, i.e., 
	${\bf 
		\Sigma}_{t-1} = {{\bf{\Sigma 
		}}_{\bf{h}}} - {{\bf{\Sigma 
		}}_{\bf{h}}} {{\bf{W}}_{t-1}} {\bf{K}}_{t-1}^{-1} 
	{{\bf{W}}_{t-1}^H}{{\bf{\Sigma 
		}}_{\bf{h}}}$. Then, by substituting (\ref{eq:L1p2}) and 
	(\ref{eq:L1p3}) into 
	(\ref{eq:L1p1}), 
	the inverse of ${\bf{K}}_t$ can be written as (\ref{eqn:K_t_inv}).
	\begin{figure*}[!b]
		\normalsize	
		\hrulefill
		\begin{align}\label{eqn:K_t_inv}
			{\bf{K}}_t^{-1} = 	\left[
			\begin{array}{cc}
				{\bf{K}}_{t-1}^{-1} + 
				\dfrac{{\bf{K}}_{t-1}^{-1}{{\bf{W}}_{t-1}^H}{{\bf{\Sigma 
						}}_{\bf{h}}} {\bf{w}}_t{\bf{w}}_t^H{{\bf{\Sigma 
						}}_{\bf{h}}} 
					{{\bf{W}}_{t-1}}{\bf{K}}_{t-1}^{-1}}{{\bf{w}}_t^H {\bf 
						\Sigma}_{t-1} {\bf{w}}_t + \sigma^2}	&  	- 
				\dfrac{{\bf{K}}_{t-1}^{-1} 
					{{\bf{W}}_{t-1}^H}{{\bf{\Sigma }}_{\bf{h}}} 
					{\bf{w}}_t}{{\bf{w}}_t^H 
					{\bf \Sigma}_{t-1} {\bf{w}}_t + \sigma^2}
				\vspace*{0.3cm}
				\\ 		
				- \dfrac{{\bf{w}}_t^H{{\bf{\Sigma }}_{\bf{h}}} {{\bf{W}}_{t-1}} 
					{\bf{K}}_{t-1}^{-1}}{{\bf{w}}_t^H {\bf \Sigma}_{t-1} 
					{\bf{w}}_t + 
					\sigma^2}
				&			\dfrac{1}{{\bf{w}}_t^H {\bf \Sigma}_{t-1} 
					{\bf{w}}_t + 
					\sigma^2}
			\end{array}\right]
		\end{align}
	\end{figure*}
	Taking ${\bf{K}}_t^{-1}$ back to the posterior kernel ${\bf \Sigma}_t$ 
	in 
	(\ref{eq:kernel}) results in 
	\begin{align}\label{eqn:Sigma_t_EFGH}
		{\bf \Sigma}_t =& {{\bf{\Sigma }}_{\bf{h}}} - {{\bf{\Sigma 
			}}_{\bf{h}}}{{\bf{W}}_t}{\bf{K}}_t^{-1}
		{ {{{\bf{W}}_t^H}} }
		{{\bf{\Sigma }}_{\bf{h}}} \notag \\=& {{\bf{\Sigma }}_{\bf{h}}} - 
		{{\bf{\Sigma 
			}}_{\bf{h}}}{{\bf{W}}_{t-1}}{\bf{K}}_{t-1}^{-1}
		{ {{{\bf{W}}_{t-1}^H}} }
		{{\bf{\Sigma }}_{\bf{h}}} + \notag \\ & \frac{1}{{\bf{w}}_t^H {\bf 
				\Sigma}_{t-1} 
			{\bf{w}}_t + \sigma^2}({\bf E}+{\bf F}+{\bf G}+{\bf H}) \notag 
		\\ = &
		{\bf \Sigma}_{t-1} + \frac{1}{{\bf{w}}_t^H {\bf \Sigma}_{t-1} 
			{\bf{w}}_t + \sigma^2}({\bf E}+{\bf F}+{\bf G}+{\bf H}), 
	\end{align}
	where
	\begin{align*}
		{\bf E} &= - {{\bf{\Sigma 
			}}_{\bf{h}}}{{\bf{W}}_{t-1}}{\bf{K}}_{t-1}^{-1}
		{ {{{\bf{W}}_{t-1}^H}} }
		{{\bf{\Sigma }}_{\bf{h}}} {\bf w}_t {\bf w}_t^H {{\bf{\Sigma 
			}}_{\bf{h}}}{{\bf{W}}_{t-1}}{\bf{K}}_{t-1}^{-1}
		{ {{{\bf{W}}_{t-1}^H}} }
		{{\bf{\Sigma }}_{\bf{h}}} \notag \\&= - ({{\bf{\Sigma }}_{\bf{h}}} 
		- 
		{{\bf{\Sigma 
			}}_{t-1}})  {\bf w}_t {\bf w}_t^H ({{\bf{\Sigma }}_{\bf{h}}} - 
		{{\bf{\Sigma 
			}}_{t-1}}), \\
		{\bf F} &=  {{\bf{\Sigma 
			}}_{\bf{h}}}{{\bf{W}}_{t-1}}{\bf{K}}_{t-1}^{-1}
		{ {{{\bf{W}}_{t-1}^H}} }
		{{\bf{\Sigma }}_{\bf{h}}} {\bf w}_t {\bf w}_t^H {{\bf{\Sigma 
			}}_{\bf{h}}} \notag \\&=  ({{\bf{\Sigma }}_{\bf{h}}} - 
		{{\bf{\Sigma 
			}}_{t-1}})  {\bf w}_t {\bf w}_t^H {{\bf{\Sigma }}_{\bf{h}}}, \\
		{\bf G} &= {{\bf{\Sigma }}_{\bf{h}}} {\bf w}_t {\bf 
			w}_t^H{{\bf{\Sigma 
			}}_{\bf{h}}}{{\bf{W}}_{t-1}}{\bf{K}}_{t-1}^{-1}
		{ {{{\bf{W}}_{t-1}^H}} }
		{{\bf{\Sigma 
			}}_{\bf{h}}} \notag \\&=  {{\bf{\Sigma }}_{\bf{h}}} {\bf w}_t 
		{\bf 
			w}_t^H({{\bf{\Sigma }}_{\bf{h}}} - 
		{{\bf{\Sigma 
			}}_{t-1}}), \\
		{\bf H} &= - {{\bf{\Sigma }}_{\bf{h}}}  {\bf w}_t {\bf w}_t^H 
		{{\bf{\Sigma 
			}}_{\bf{h}}}.
	\end{align*}
	It is straightforward to verify that the sum of ${\bf E}$, ${\bf F}$, ${\bf 
		G}$, and ${\bf 
		H}$ is exactly
	$-{{\bf{\Sigma }}_{t-1}}  {\bf w}_t {\bf w}_t^H {{\bf{\Sigma 
		}}_{t-1}}$. By substituting ${\bf E}$, ${\bf F}$, ${\bf 
		G}$, and ${\bf 
		H}$ into (\ref{eqn:Sigma_t_EFGH}), we have
	\begin{align}
		{\bf{\Sigma}}_t ={\bf{\Sigma}}_{t-1} - \frac{{\bf{\Sigma}}_{t-1}  
			{\bf{w}}_{t}
			{\bf{w}}_{t}^H{\bf{\Sigma}}_{t-1}}{{\bf{w}}_{t}^H{\bf{\Sigma}}_{t-1}{\bf{w}}_{t}
			+ \sigma^2}, \end{align}
	which completes the proof. 
	
	\section{Proof of Theorem~\ref{theorem:2}}\label{appendix:thm2}
	\textbf{Theorem \ref{theorem:2}} can be proved by the contradiction method. 
	Suppose there exists an index $k$ such that $n_k \ge p_k + 1$. Due to the 
	constraint 
	that $\sum_{k=1}^K n_k = \sum_{k=1}^K p _k = Q$, there must exists $k'$ 
	such that $k'\neq k$  and $n_{k'} < p_{k'}$, otherwise $\sum_{k=1}^K n_k$ 
	will 
	be larger than 
	$\sum_{k=1}^K p _k$. Note that $p_{k'}$ is non-zero since $p_{k'} > 
	n_{k'} \ge 0$. In this context, the water level can be expressed as $\beta 
	= 
	p_{k'} + \frac{\sigma^2}{\lambda_{k'}}$ and the $k'$-th ice level 
	$\frac{\sigma^2}{\lambda_{k'}^Q}$ should be smaller than $\beta$ because 
	\begin{align}\label{eq:thm2.1}
		\frac{\sigma^2}{\lambda_{k'}^Q} = n_{k'} + 
		\frac{\sigma^2}{\lambda_{k'}} < 
		p_{k'} + \frac{\sigma^2}{\lambda_{k'}} = \beta. 
	\end{align}
	Then, consider the $k$-th ice level $\frac{\sigma^2}{\lambda_{k}^Q}$. Since 
	$n_k \ge p_k + 1$, we have the following inequality:
	\begin{align}\label{eq:thm2.2}
		\frac{\sigma^2}{\lambda_{k}^Q} &= n_{k} + \frac{\sigma^2}{\lambda_{k}} 
		\ge 
		p_{k} + \frac{\sigma^2}{\lambda_{k}} + 1 = \left(\beta - 
		\frac{\sigma^2}{\lambda_{k}}\right)^{+} + 
		\frac{\sigma^2}{\lambda_{k}} + 1 \notag \\
		& \overset{(a)}{\ge} \beta + 1,
	\end{align}
	where the inequality (a) holds because $\left(\beta - 
	\frac{\sigma^2}{\lambda_{k}}\right)^{+} \ge \beta - 
	\frac{\sigma^2}{\lambda_{k}}$. Combining equations \eqref{eq:thm2.1} and 
	\eqref{eq:thm2.2}, we arrive at
	\begin{align}\label{eq:thm2.3}
		\frac{\sigma^2}{\lambda_{k}^Q} > \frac{\sigma^2}{\lambda_{k'}^Q} + 1.
	\end{align}
	We introduce the following \textbf{Lemma~2.1} to evaluate the 
	inequality \eqref{eq:thm2.3}. 
	
	\textbf{Lemma 2.1:}
	For the $k$-th eigenvalue $\lambda_{k}^t$ obtained at the $t$-th timeslot, 
	if $n_k^t > 0$, we have the inequality
	\begin{align}\label{eq:l7}
		\frac{\sigma^2}{\lambda_{k'}^t} + 1 \ge 
		\frac{\sigma^2}{\lambda_{k}^t},  	
	\end{align}
	hold for all $k'\in\{1,2,\cdots, K\}$.
	
	\begin{IEEEproof}
		Since $n_k^t > 0$, the $k$-th eigenvector is selected by the 
		ice-filling 
		algorithm at least once. 
		Let $t' < t$ denote the latest timeslot before $t$ when the $k$-th 
		eigenvector is selected for pilot transmission such that $n_k^{t'} = 
		n_k^t 
		- 
		1$ and  $n_k^{t' + 1} = n_k^t$.
		
		Suppose there exists $k'$ such that $\frac{\sigma^2}{\lambda_{k'}^t} < 
		\frac{\sigma^2}{\lambda_{k}^t} - 1$, then we have 
		\begin{align}\label{eq:pl7}
			\frac{\sigma^2}{\lambda_{k'}^{t'}} \overset{(b)}{\le} 
			\frac{\sigma^2}{\lambda_{k'}^{t}} \overset{(c)}{<} 
			\frac{\sigma^2}{\lambda_k} + n_k^t - 1 = 
			\frac{\sigma^2}{\lambda_k} + 
			n_k^{t'} =  \frac{\sigma^2}{\lambda_{k}^{t'}},
		\end{align}
		where (b) holds because the ice-level 
		$\frac{\sigma^2}{\lambda_{k'}^{t}}$ 
		is 
		non-decreasing w.r.t the timeslot $t$, and (c) holds given 
		that $\frac{\sigma^2}{\lambda_k} + n_k^t = 
		\frac{\sigma^2}{\lambda_{k}^{t}}$.
		The inequality $\frac{\sigma^2}{\lambda_{k'}^{t'}} < 
		\frac{\sigma^2}{\lambda_{k}^{t'}}$ in \eqref{eq:pl7} contradicts the 
		eigenvector selection principle of the ice-filling algorithm that 
		$\lambda_{k}^{t'}$ should 
		be the largest over the eigenvalue set $\{\lambda_{1}^{t'}, 
		\lambda_{2}^{t'}, 
		\cdots, \lambda_{K}^{t'}\}$. Therefore, 
		$\frac{\sigma^2}{\lambda_{k'}^t}$ 
		should be no less than  $\frac{\sigma^2}{\lambda_{k}^t} - 1$ for all 
		$k'$, 
		which 
		completes the proof. 
	\end{IEEEproof}
	
	Comparing \eqref{eq:thm2.3} and \eqref{eq:l7}, it is clear that 
	\eqref{eq:thm2.3} contradicts \textbf{Lemma 2.1} given that 
	$n_k \ge p_k + 1 > 0$. Therefore, 
	$n_k$ 
	should be smaller 
	than $p_k + 1$. 
	
	We adopt the same method to prove that $n_k > p_k - 1$. 
	Specifically, suppose there is $k$ such that $n_k \le p_k - 1$. Then there 
	must 
	exist $k'\neq k$ such that $n_{k'} > p_{k'}$. From the inequality $n_k \le 
	p_k 
	- 1$, we know that the ice level 
	$\frac{\sigma^2}{\lambda_k^Q}$ satisfies the inequality
	\begin{align}\label{eq:thm2.4}
		\frac{\sigma^2}{\lambda_k^Q} = n_k + \frac{\sigma^2}{\lambda_k} \le p_k 
		+ 
		\frac{\sigma^2}{\lambda_k}
		- 1 \overset{(b)}{=} \beta - 1,
	\end{align}
	where (b) arises because $p_k$ is greater than 0. 
	On the other hand, from the inequality $n_{k'} > p_{k'}$, we get 
	\begin{align}\label{eq:thm2.5}
		\frac{\sigma^2}{\lambda_{k'}^Q} = n_{k'} + 
		\frac{\sigma^2}{\lambda_{k'}} 
		> 
		p_{k'} 
		+ 
		\frac{\sigma^2}{\lambda_{k'}} \ge \beta.
	\end{align}
	Combining \eqref{eq:thm2.4} and \eqref{eq:thm2.5}, we get 
	\begin{align}\label{eq:thm2.6}
		\frac{\sigma^2}{\lambda_{k'}^Q} \ge \frac{\sigma^2}{\lambda_{k}^Q} + 1,
	\end{align}
	which contradicts \textbf{Lemma 2.1} because $n_{k'} > p_{k'} 
	\ge 0$. As a result, $n_{k}$ should be larger than  $p_{k} - 1$, and 
	thereby we 
	get 
	\begin{align}
		|n_k - p_k| < 1,
	\end{align}
	which completes the proof.


	\section{Proof of Lemma~\ref{lemma:2}}\label{appendix:2}
	
	Substituting $\mb{W} = \mb{U}_K\mb{P}$ and  ${\bf 
		\Sigma_h} = \mb{U}_K{\bf \Lambda}_K\mb{U}_K^H$ into the definition 
	of 
	${{\bf{\Sigma }}}_{\mb{h}|\mb{y}}$, we get
	\begin{align}
		&{{\bf{\Sigma }}}_{\mb{h}|\mb{y}} = {{\bf{\Sigma 
			}}_{\bf{h}}} -
		{{\bf{\Sigma 
			}}_{\bf{h}}}{{\bf{W}}}
		{\left( {\bf W}^{H}{\bf{\Sigma 
			}}_{\bf{h}}{\bf W} + \sigma^2 {\bf I}_Q \right)^{ - 
				1}}{ 
			{{{\bf{W}}^H}} }
		{{\bf{\Sigma }}_{\bf{h}}} \notag \\
		=& \mb{U}_K{\bf \Lambda}_K\mb{U}_K^H - \mb{U}_K{\bf 
			\Lambda}_K\mb{U}_K^H 
		\mb{U}_K\mb{P}(\mb{P}^H\mb{U}_K^H 
		\mb{U}_K{\bf \Lambda}_K\mb{U}_K^H\mb{U}_K\mb{P} \notag \\  
		&+ \sigma^2 
		\mb{I}_Q 
		)^{-1}\mb{P}^H\mb{U}_K^H\mb{U}_K{\bf \Lambda}_K\mb{U}_K^H 
		\notag 
		\\
		=& \mb{U}_K \underbrace{\left({\bf \Lambda}_K - {\bf \Lambda}_K 
			\mb{P}\left(\mb{P}^H{\bf \Lambda}_K\mb{P} + \sigma^2 
			\mb{I}_Q 
			\right)^{-1}\mb{P}^H{\bf \Lambda}_K  \right)  }_{{\bf 
				\Lambda}_{\mb{h}|\mb{y}}}
		\mb{U}_K^H.
	\end{align}
	Note that matrices ${\bf \Lambda}_K$ and $\mb{P}$ are both 
	diagonal. Hence, it can be directly proved that
	\begin{align}
		{\bf 
			\Lambda}_{\mb{h}|\mb{y}} = 
		{\rm diag}\left(\frac{\lambda_1\sigma^2}{p_1\lambda_1 + 
			\sigma^2}, \cdots, \frac{\lambda_K\sigma^2}{p_K\lambda_K + 
			\sigma^2}\right).
	\end{align}
	Therefore, given that $\mb{U}_K^H\mb{U}_K = 
	\mb{I}_K$, the MSE achieved by water-filling can be derived as 
	\begin{align}
		\mathsf{Tr}\left({{\bf{\Sigma }}}_{\mb{h}|\mb{y}}\right) =  
		\mathsf{Tr}\left({\bf 
			\Lambda}_{\mb{h}|\mb{y}}\right) = \sum_{k = 1}^{K} \frac{\lambda_k 
			\sigma^2}{p_k 
			\lambda_k + \sigma^2}.
	\end{align} 
	Particularly, as $Q \rightarrow +\infty$, the water-filling principle 
	asymptotically allocates equal power $Q/K$ to all 
	eigenvectors~\cite{Tse2005WC}, rendering the 
	estimation error decays at a rate of $\mathcal{O}\left(\sum_{k = 1}^K 
	\frac{K\sigma^2}{Q}\right) = \mathcal{O}\left(K^2Q^{-1}\right)$. This 
	completes the 
	proof. 
	
	\section{Proof of Lemma~\ref{lemma:3}}\label{appendix:3}
	We first express the observation matrix designed by the ice-filling 
	algorithm 
	as 
	$\mb{W} = \mb{U}_K \mb{S}$. Here, $\mb{S} \in \{0, 
	1\}^{K\times Q}$ represents a switch matrix, each column of which contains
	one non-zero entry, i.e., $\mb{S}(k,q) \in \{0, 1\}$ and $\sum_{k = 
		1}^{K}\mb{S}(k,q) = 1$. In particular, $\mb{S}(k,q) = 1$ implies that 
		the 
	$k$-th eigenvector, $\mb{u}_k$, is 
	assigned to the $q$-th observation vector. Based on this definition, we can 
	prove 
	that
	\begin{align}\label{eq:N}
		\mb{S}\mb{S}^H = \mb{N} = {\rm diag}(n_1, n_2, \cdots, n_K),
	\end{align} 
	given that $n_k$ ($k\in\{1,2,\cdots, K\}$) stands for the number of times 
	that 
	the 
	eigenvector $\mb{u}_k$ is assigned to $\mb{W}$ by 
	\textbf{Algorithm~\ref{alg:IF}}. 
	Then, 
	by substituting $\mb{W} = 
	\mb{U}_K\mb{S}$ and  ${\bf 
		\Sigma_h} = \mb{U}_K{\bf \Lambda}_K\mb{U}_K^H$ into \eqref{eq:kernel}, 
	we can express ${{\bf{\Sigma }}}_{\mb{h}|\mb{y}}$ as
	\begin{align}
		{{\bf{\Sigma }}}_{\mb{h}|\mb{y}} = \mb{U}_K {\bf 
			\Lambda}_{\mb{h}|\mb{y}}
		\mb{U}_K^H,
	\end{align}
	where ${\bf 
		\Lambda}_{\mb{h}|\mb{y}} = {\bf 
		\Lambda}_K - {\bf \Lambda}_K 
	\mb{S}\left(\mb{S}^H{\bf \Lambda}_K\mb{S} + \sigma^2 
	\mb{I}_Q 
	\right)^{-1}\mb{S}^H{\bf \Lambda}_K$. 
	Moreover, the Sherman-Morrison-Woodbury formula~\cite{Meyer2000Matrix} 
	\begin{align}\label{eq:Woodbury}
		(\mb{A} + 
		\mb{X}\mb{R}\mb{Y})^{-1} = 
		\mb{A}^{-1} 
		- 
		\mb{A}^{-1}\mb{X}(\mb{R}^{-1}+\mb{Y}\mb{A}^{-1}\mb{X})^{-1}\mb{Y}\mb{A}^{-1}\end{align}
	allows us to rewrite ${\bf \Lambda}_{\mb{h}|\mb{y}}$ as 
	\begin{align} \label{eq:diagonal}
		\notag
		&{\bf \Lambda}_{\mb{h}|\mb{y}} = {\bf 
			\Lambda}_K - \frac{1}{\sigma^2} {\bf 
			\Lambda}_K \mb{S}\mb{S}^H{\bf 
			\Lambda}_K \\ &~~~~~~+ \frac{1}{\sigma^4} {\bf 
			\Lambda}_K \mb{S}\mb{S}^H
		\left({\bf \Lambda}_K^{-1} + 
		\frac{1}{\sigma^2}\mb{S}\mb{S}^H\right)^{-1}\mb{S}\mb{S}^H
		{\bf 
			\Lambda}_K \notag\\
		&\overset{(a)}{=} {\bf 
			\Lambda}_K \!-\! \frac{1}{\sigma^2} {\bf 
			\Lambda}_K  \mb{N} {\bf 
			\Lambda}_K \!+\! \frac{1}{\sigma^4} {\bf 
			\Lambda}_K  \mb{N} 
		\left({\bf \Lambda}_K^{-1} \!+\!
		\frac{1}{\sigma^2} \mb{N} \right)^{\!\!-1} \!\! \mb{N}
		{\bf \Lambda}_K, \notag \\
		&\overset{(b)}{=}  
		{\rm diag}\left(\frac{\lambda_1\sigma^2}{n_1\lambda_1 + 
			\sigma^2}, \cdots, \frac{\lambda_K\sigma^2}{n_K\lambda_K + 
			\sigma^2}\right),
	\end{align}
	where $(a)$ holds since $\mb{S}\mb{S}^H = \mb{N}$, and the equation $(b)$ 
	comes 
	directly from the fact the matrices $\mb{\Lambda}_K$ and $\mb{N}$ are both 
	diagonal. 
	%
	In conclusion, the MSE achieved by the ice-filling algorithm is  
	$\mathsf{Tr}({\bf 
		\Lambda}_{\mb{h}|\mb{y}}) =  \sum_{k = 1}^{K} \frac{\lambda_k 
		\sigma^2}{n_k 
		\lambda_k + \sigma^2}$. Moreover, since $|n_k - p_k| < 1$, the 
		asymptotic 
	estimation error is naturally $\mathcal{O}(K^2Q^{-1})$ as 
	$Q\rightarrow +\infty$, which completes the proof.
	
	\section{Proof of Theorem~\ref{thm:3}}\label{appendix:MSE_thm}
	We use the \textbf{Theorem~\ref{theorem:2}} to upper-bound the asymptotic 
	MSE 
	difference as $Q \rightarrow +\infty$, 
	i.e., 
	\begin{align}
		|\delta_{\rm wf} - \delta_{\rm if}| 
		&= \left| \sum_{k = 1}^{K} 
		\frac{\lambda_k 
			\sigma^2}{p_k 
			\lambda_k + \sigma^2} - \sum_{k = 1}^{K} 
		\frac{\lambda_k 
			\sigma^2}{n_k 
			\lambda_k + \sigma^2}\right| \notag \\ 
		&\le  
		\sum_{k = 1}^{K} \left| 
		\frac{\lambda_k 
			\sigma^2}{p_k 
			\lambda_k + \sigma^2} - 
		\frac{\lambda_k 
			\sigma^2}{n_k 
			\lambda_k + \sigma^2}\right| \notag \\
		&=
		\sum_{k = 1}^{K} \left| 
		\frac{\lambda_k^2 
			\sigma^2(p_k - n_k)}{(p_k 
			\lambda_k + \sigma^2)(n_k 
			\lambda_k + \sigma^2)}\right| \notag \\
		&\overset{(a)}{<}\sum_{k = 1}^{K} \left| 
		\frac{\lambda_k^2 
			\sigma^2}{(p_k 
			\lambda_k + \sigma^2)((p_k - 1) 
			\lambda_k + \sigma^2)}\right|,
	\end{align}
	where (a) holds due to the facts that $|n_k - p_k| < 1$ and $n_k > p_k - 1 
	> 0$ 
	when $Q\rightarrow 
	+\infty$. Moreover, the 
	water-filling principle tends to allocate equal power to all channels, 
	i.e., 
	$p_k\rightarrow Q/K$  for large $Q$~\cite{Tse2005WC}. Thereafter, the 
	asymptotic upper-bound of 
	the MSE 
	difference becomes
	\begin{align}
		|\delta_{\rm wf} - \delta_{\rm if}| < 
		\sum_{k=1}^K\frac{K^2\sigma^2}{Q^2} = 
		\mathcal{O}(K^3Q^{-2}). 
	\end{align}
	\section{Proof of Lemma~\ref{lemma:rand_MM}}\label{appendix:MM_MSE}
	Let ${\bf \Sigma_h} = \mb{U}_K{\bf \Lambda}_K\mb{U}_K^H$ denote the 
	eigenvalue 
	decomposition of the rank-$K$ kernel ${{\bf{\Sigma }}_{\bf{h}}}$, wherein 
	$\mb{U}_K\in{\mathbb C}^{M\times K}$ and $\mb{\Lambda}_K\in{\mathbb 
	C}^{M\times 
		M}$. According to the extension of the Sherman-Morrison-Woodbury 
		formula in 
	\eqref{eq:Woodbury}, 
	the MSE achieved by the proposed algorithms can be rewritten as
	\begin{align}\label{eqn:45}
		&\mathsf{E}\left(\|{\bm{\mu}_{\mb{h}|\mb{y}}} - 
		\mb{h}\|^2_2\right) = \mathsf{Tr}\left({{\bf{\Sigma 
		}}}_{\mb{h}|\mb{y}}\right) \notag \\=&
		\mathsf{Tr} \left({ {{\bf{U}}_K}{\left( {{\bf{\Lambda }}_K^{ - 1} + {1 
						\over {{\sigma 
						^2}}}{\bf{U}}_K^H{\bf{W}}{{\bf{W}}^H}{{\bf{U}}_K}} 
						\right)^{ 
					- 1}}{\bf{U}}_K^H }\right).
	\end{align}
	For the randomly generated observation matrix $\bf W$ in both the 
	amplitude-and-phase 
	controllable and phase-only controllable cases, following the asymptotic 
	orthogonality principle,
	we have 
	${\bf{W}}{{\bf{W}}^H} \approx {Q \over M}{{\bf{I}}_M}$ when $Q$ is 
	sufficiently 
	large. In this 
	case, the posterior kernel can be approximated by
	\begin{align}\label{eqn:46}
		&{{\bf{\Sigma }}}_{\mb{h}|\mb{y}} \approx {{\bf{U}}_K}{\left( 
		{{\bf{\Lambda 
				}}_K^{ - 1} + {1 \over {{\sigma ^2}}}{Q \over 
				M}{\bf{U}}_K^H{{\bf{U}}_K}} 
			\right)^{ - 1}}{\bf{U}}_K^H = \notag \\ & 
		{{\bf{U}}_K}{\rm{diag}}\left( {{1 \over {{1 \over {{\lambda _1}}} + {Q 
						\over {M{\sigma ^2}}}}},{1 \over {{1 \over {{\lambda 
						_2}}} 
					+ {Q \over 
						{M{\sigma ^2}}}}}, \cdots ,{1 \over {{1 \over {{\lambda 
								_K}}} + {Q \over 
						{M{\sigma ^2}}}}}} \right){\bf{U}}_K^H.
	\end{align}
	Therefore, by substituting (\ref{eqn:46}) 
	into (\ref{eqn:45}), the MSE $\delta_{\rm rnd}$ is given by
	\begin{align}
		\delta_{\rm rnd} \approx	\sum\limits_{k = 1}^K {{1 \over {{1 \over 
						{{\lambda 
								_k}}} + {Q \over {M{\sigma ^2}}}}}} = 
		\sum\limits_{k = 1}^K {{{{\lambda _k}{\sigma ^2}} 
				\over {{Q \over M}{\lambda _k} + {\sigma ^2}}}} ,
	\end{align}
	which completes the proof.
	
	\section{Proof of Lemma \ref{lemma:4}}\label{appendix:4}
	Given a prior kernel $\bf \Sigma$ as the input of the 
	proposed estimator, the squared error of channel estimation can be derived 
	as 
	\begin{align}
		&{{{\left\| {{{\bm{\mu }}}_{\mb{h}|\mb{y}} - {\bf{h}}} \right\|}^2_2}}
		= \notag
		{\left\| {{\bf{\Sigma W}}{{\left( {{{\bf{W}}^H}{\bf{\Sigma W}} + 
							{\sigma ^2}{{\bf{I}}_Q}} \right)}^{ - 1}}{\bf{y}} - 
				{\bf{h}}} 
			\right\|^2_2}
		\\ 
		\notag
		&\stackrel{(a)}{=}  {\Big \|} 
		\left( {{\bf{\Sigma W}}{{\left( {{{\bf{W}}^H}{\bf{\Sigma W}} + {\sigma 
							^2}{\bf I }_Q} \right)}^{ - 1}}{{\bf{W}}^H} - 
			{{\bf{I}}_M}} 
		\right){{\bf{h}}} \\ & \notag ~~~~~~~~+ 
		{\bf{\Sigma W}}{\left( {{{\bf{W}}^H}{\bf{\Sigma W}} + {\sigma ^2}{\bf I 
				}_Q} \right)^{ - 1}}{\bf{z}}
		{\Big \|}^2_2
		\\
		& \stackrel{(b)}{=} {\left\| \left( {{{\bf{\Pi }}^H}{{\bf{W}}^H} - 
				{{\bf{I}}_M}} \right){\bf{h}} + {{\bf{\Pi }}^H}{\bf{z}} 
			\right\|^2_2},
	\end{align}
	where $(a)$ holds since ${\bf y} = {\bf W}^{H}{\bf h} + {\bf z}$ and $(b)$ 
	holds by defining ${\bf{\Pi }} = {\left( {{{\bf{W}}^H}{\bf{\Sigma W}} + 
			{\sigma ^2}{\bf I }_Q} \right)^{ - 1}}{{\bf{W}}^H}{\bf{\Sigma }}$, 
	which 
	can 
	be viewed as a matrix function with respect to the artificial kernel $\bf 
	\Sigma$. 
	Next, recalling properties ${\bf z} \sim \mathcal{C} 
	\mathcal{N}\!\left({\bf 0}_{Q}, \sigma^2{\bf I}_Q \right)$ and ${\mathsf 
		E}\left({\bf h}{\bf h}^{\rm 
		H}\right) = {\bf{\Sigma }}_{{\bf h}}$, the MSE ${\hat \delta}$ under 
	imperfect kernel can be derived as
	\begin{align}
		{\hat \delta}
		= \mathsf{Tr}\left(\left( {{{\bf{\Pi }}^H}{{\bf{W}}^H} - {{\bf{I}}_M}} 
		\right)\!{{\bf{\Sigma }}_{\bf{h}}}\!\left( {{\bf{W\Pi }} - 
		{{\bf{I}}_M}} 
		\right)\right)  + {\sigma ^2} 	\mathsf{Tr}\!\left({{\bf{\Pi 
		}}^H}{\bf{\Pi 
		}} \right),
	\end{align}
	where the equality holds since ${\left\| {\bf{x}} \right\|^2_2} = 
	{\mathsf{Tr}}\left( {{\bf{x}}{{\bf{x}}^{\rm H}}} \right)$ for any vector 
	$\bf x$, which completes the proof.

	\section{Proof of Lemma 
		\ref{lemma:useful_lemma_2}}\label{appendix:useful_lemma_2}
	Comparing (\ref{eqn:useful_hat_delta}) and (\ref{eqn:nonideal_MSE}), we 
	only 
	need to prove that ${\bf{W\Pi }} = {\bf{\Omega }}{{{\bf{\hat \Sigma 
		}}}_{\bf{h}}}$ and ${{\bf{\Pi }}^H}{\bf{\Pi }} = {{{\bf{\hat \Sigma 
		}}}_{\bf{h}}}{\bf{\Xi }}{{{\bf{\hat \Sigma }}}_{\bf{h}}}$ in the case 
		when 
	${\bf{\Sigma }}={\bf{\hat \Sigma }}_{{\bf h}}$. 
	Due to the positive semi-definition of ${\bf{ 
			\Sigma }}_{{\bf h}}$, the inversion of ${\bf{\hat \Sigma }}_{{\bf 
			h}} = 
	{\bf{ 
			\Sigma }}_{{\bf 
			h}}+ \sigma^2_{\bf h}{\bf I}_M$ exists. 
	Thus, by applying the Sherman-Morrison-Woodbury formula in 
	\eqref{eq:Woodbury} 
	for the term ${\left( {{{\bf{W}}^H}{{{\bf{\hat \Sigma }}}_{\bf{h}}}{\bf{W}} 
	+ 
			{\sigma ^2}{{\bf{I}}_Q}} \right)^{ - 1}}$, ${\bf{W\Pi }}$ can be 
	reorganized as
	\begin{align}\label{eqn:useful_lemma_2_eq1}
		& {\bf{W}}{{\bf{\Pi }}} = {\bf{W}}{\left( {{{\bf{W}}^H}{{{\bf{\hat 
		\Sigma 
					}}}_{\bf{h}}}{\bf{W}} + {\sigma ^2}{{\bf{I}}_Q}} \right)^{ 
					- 
				1}}{{\bf{W}}^H}{{{\bf{\hat \Sigma }}}_{\bf{h}}} \notag \\ 
		&  = {\bf{W}}\left( {{1 \over {{\sigma ^2}}}{{\bf{I}}_Q} - {1 \over 
				{{\sigma ^4}}}{{\bf{W}}^H}{{\left( {{\bf{\hat \Sigma 
				}}_{\bf{h}}^{ - 1} 
						\!+\! {1 \over {{\sigma ^2}}}{\bf{W}}{{\bf{W}}^H}} 
						\right)}^{\! - 
					1}}{\bf{W}}} \right){{\bf{W}}^H}{{{\bf{\hat \Sigma 
					}}}_{\bf{h}}}  \notag \\ 
		&  = \left( {{{{\bf{W}}{{\bf{W}}^H}} \over {{\sigma ^2}}} \!-\! 
			{{{\bf{W}}{{\bf{W}}^H}} \over {{\sigma ^4}}}{{\left( {{\bf{\hat 
			\Sigma 
						}}_{\bf{h}}^{ \!- 1} \!+\! {{{\bf{W}}{{\bf{W}}^H}} 
						\over {{\sigma ^2}}}} 
					\right)}^{ - 1}}{\bf{W}}{{\bf{W}}^H}} \right){{{\bf{\hat 
					\Sigma 
			}}}_{\bf{h}}},
	\end{align}
	which is equal to ${\bm \Omega}{{\bf{\hat \Sigma }}}_{\bf{h}}$ and 
	obviously 
	the is a function of ${\bf W}{\bf W}^H$. 
	Using the same matrix techniques as 
	(\ref{eqn:useful_lemma_2_eq1}), we can prove that 
	\begin{align}
		\notag
		&{{\bf{\Pi }}^H}{\bf{\Pi }} = {{{\bf{\hat \Sigma 
			}}}_{\bf{h}}}{\bf{W}}{\left( {{{\bf{W}}^H}{{{\bf{\hat \Sigma 
					}}}_{\bf{h}}}{\bf{W}} + {\sigma ^2}{{\bf{I}}_Q}} \right)^{ 
					- 
				2}}{{\bf{W}}^H}{{{\bf{\hat \Sigma }}}_{\bf{h}}}  \\ \notag
		=& {{{\bf{\hat \Sigma }}}_{\bf{h}}}{\bf{W}}{\left( {{1 \over {{\sigma 
							^2}}}{{\bf{I}}_Q} \!-\! {1 \over {{\sigma 
							^4}}}{{\bf{W}}^H}{{\left( 
						{{\bf{\hat \Sigma }}_{\bf{h}}^{- 1} \!+\! 
						{{{\bf{W}}{{\bf{W}}^H}} \over 
								{{\sigma ^2}}}} \right)}^{ \!\!- 1}}{\bf{W}}} 
			\right)^{\!\!2}}{{\bf{W}}^H}{{{\bf{\hat \Sigma }}}_{\bf{h}}} \\ 
			\notag
		=& {{{\bf{\hat \Sigma }}}_{\bf{h}}} {\Bigg (}{{{\bf{W}}{{\bf{W}}^H}} 
		\over 
			{{\sigma ^4}}} - 2{{{\bf{W}}{{\bf{W}}^H}} \over {{\sigma 
			^6}}}{\left( 
			{{\bf{\hat \Sigma }}_{\bf{h}}^{ - 1} + {{{\bf{W}}{{\bf{W}}^H}} 
			\over 
					{{\sigma ^2}}}} \right)^{ - 1}}{\bf{W}}{{\bf{W}}^H} + \\ 
		&{{{\bf{W}}{{\bf{W}}^H}} \over {{\sigma ^8}}} \left({\left( {{\bf{\hat 
						\Sigma 
				}}_{\bf{h}}^{ - 1} + {{{\bf{W}}{{\bf{W}}^H}} \over {{\sigma 
				^2}}}} 
			\right)^{ - 1}}{\bf{W}}{{\bf{W}}^H}\right)^2 { \Bigg  )}{{{\bf{\hat 
						\Sigma 
			}}}_{\bf{h}}},
	\end{align}
	which is equal to ${{{\bf{\hat \Sigma }}}_{\bf{h}}}{\bf{\Xi }}{{{\bf{\hat 
					\Sigma }}}_{\bf{h}}}$ and obviously is a function of ${\bf 
		W}{\bf W}^H$. This 
	completes the proof.
	
	\section{Proof of Lemma 
		\ref{lemma:MSE_under_imp_kernel}}\label{appendix:MSE_under_imp_kernel}
	Let ${\bf \Sigma_h} = \mb{U}{\bf \Lambda}\mb{U}^H$ denote the complete 
	eigenvalue decomposition of the kernel ${{\bf{\Sigma }}_{\bf{h}}}$, wherein 
	$\mb{U}\in{\mathbb C}^{M\times M}$ and $\mb{\Lambda} = 
	\text{diag}\{\lambda_1, \cdots, \lambda_M\}$ with $\lambda_m = 0$ for 
	$K<m\le 
	M$. 
	According to the analyses in Appendices \ref{appendix:2}, \ref{appendix:3}, 
	and 
	\ref{appendix:MM_MSE}, the term ${\bf W}{\bf W}^H$ for the water-filling 
	estimator, ice-filling estimator, and random estimator can be respectively 
	written as
	\begin{subequations}\label{eqn:WWH}
		\begin{align}
			& {{\bf{W}}_{{\rm{wf}}}}{\bf{W}}_{{\rm{wf}}}^H = {\bf{U\hat 
					P}}{{{\bf{\hat P}}}^H}{{\bf{U}}^H},  \\ 
			& {{\bf{W}}_{{\rm{if}}}}{\bf{W}}_{{\rm{if}}}^H = {\bf{U\hat 
					N}}{{\bf{U}}^H},  \\ 
			& {{\bf{W}}_{{\rm{rnd}}}}{\bf{W}}_{{\rm{rnd}}}^H \approx {Q \over 
				M}{{\bf{I}}_M} = {\bf{U}}\left( {{Q \over M}{{\bf{I}}_M}} 
			\right){{\bf{U}}^H}.
		\end{align}
	\end{subequations}
	where ${\hat{\bf P}}={\rm diag}\left(\sqrt{{\hat p}_1},\cdots,\sqrt{{\hat 
			p}_M}\right)$ is the power allocated to the $M$ eigenvectors of 
	$\hat{\bf 
		\Sigma}_{\bf h}$ and ${\hat{\bf N}}={\rm diag}\left({{\hat 
			n}_1},\cdots,{{\hat 
			n}_M}\right)$ is the pilot-reuse-frequency matrix, as defined in 
	\eqref{eq:N}. 
	It is clear from (\ref{eqn:WWH}) that the terms ${\bf W}{\bf 
		W}^H$ in the three cases share a general form ${\bf W}{\bf 
		W}^H = {\bf{U}}{\bm \Psi}{\bf{U}}^H$ 
	wherein ${\bm \Psi}={\rm diag}\left(\psi_1,\cdots,\psi_M\right)$ is a 
	diagonal 
	matrix. Therefore, the proof of {\bf Lemma 
		\ref{lemma:MSE_under_imp_kernel}} 
	can be obtained by replacing ${\bm \Psi}$ with the corresponding terms 
	${\bf{\hat P}}{{{\bf{\hat P}}}^H}$, ${\bf{\hat N}}$, and 
	${{Q \over M}{{\bf{I}}_M}}$,respectively. It allows us focus on the general 
	proof based on the common expression ${\bf{U}}{\bm \Psi}{\bf{U}}^H$, which 
	is 
	given as 
	follows.
	
	Recalling the useful expression in {\bf Lemma \ref{lemma:useful_lemma_2}}, 
	by 
	substituting ${\bf W}{\bf W}^H={\bf{U}}{\bm \Psi}{\bf{U}}^H$, ${\bf 
	\Sigma_h} = 
	\mb{U}{\bf \Lambda}\mb{U}^H$, and ${\bf{\hat \Sigma }}_{{\bf h}}={\bf{ 
	\Sigma 
	}}_{{\bf h}}+ \sigma^2_{\bf h}{\bf I}_M$ into $\bm \Omega$ in 
	(\ref{eqn:Omega}) 
	and $\bm \Xi$ in (\ref{eqn:Xi}), we arrive at
	\begin{align}
		\notag
		&{\bm \Omega}{{{\bf{\hat \Sigma }}}_{\bf{h}}} = {\bf{U}}\left( 
		{{{\bf{\Psi 
				}} \over {{\sigma ^2}}} - {{\bf{\Psi }} \over {{\sigma 
				^4}}}{{\left( 
					{{{\left( {{\bf{\Lambda }} + \sigma 
					_{\bf{h}}^2{{\bf{I}}_M}} \right)}^{ - 
								1}} + {{\bf{\Psi }} \over {{\sigma ^2}}}} 
								\right)}^{ - 1}}{\bf{\Psi }}} 
		\right){{\bf \Lambda}\mb{U}^H} \\ \notag
		&= {\bf{U}}{\rm{diag}}\!\left( {{{{\psi _1}\left( {{\lambda _1} \!+\! 
						\sigma _{\bf{h}}^2} \right)} \over {{\sigma ^2} \!+\! 
						{\psi _1}\left( 
					{{\lambda _1} \!+\! \sigma _{\bf{h}}^2} \right)}}, \cdots 
					,{{{\psi 
						_M}\left( {{\lambda _M} \!+\! \sigma _{\bf{h}}^2} 
						\right)} \over {{\sigma 
						^2} \!+\! {\psi _M}\left( {{\lambda _M} \!+\! \sigma 
						_{\bf{h}}^2} 
					\right)}}} \right){{\bf{U}}^H} \\
		& = {\bf{U}}{\bm \Lambda}_{\bm \Omega}{{\bf{U}}^H},
	\end{align}
	\begin{align}
		\notag
		&{{{\bf{\hat \Sigma }}}_{\bf{h}}}{\bf{\Xi }}{{{\bf{\hat \Sigma 
			}}}_{\bf{h}}} =  {\bf{U}}\left( {{\bf{\Lambda }} + \sigma 
			_{\bf{h}}^2{{\bf{I}}_M}} \right)\times \\ \notag &
		{\Bigg (}{{\bf{\Psi }} \over {{\sigma ^4}}} - 2{{\bf{\Psi }} \over 
		{{\sigma 
					^6}}}{\left( {{{\left( {{\bf{\Lambda }} + \sigma 
					_{\bf{h}}^2{{\bf{I}}_L}} 
						\right)}^{ - 1}} + {{\bf{\Psi }} \over {{\sigma ^2}}}} 
						\right)^{ - 
				1}}{\bf{\Psi }} + \\  \notag &
		{{\bf{\Psi }} \over {{\sigma ^8}}}{\left( {{{\left( {{{\left( 
		{{\bf{\Lambda 
										}} + \sigma _{\bf{h}}^2{{\bf{I}}_L}} 
										\right)}^{ - 1}} + {{\bf{\Psi }} \over 
								{{\sigma ^2}}}} \right)}^{ - 1}}{\bf{\Psi }}} 
								\right)^2} {\Bigg )}\left( 
		{{\bf{\Lambda }} + \sigma _{\bf{h}}^2{{\bf{I}}_M}} \right){{\bf{U}}^H} 
		= \\ 
		\notag &
		{\bf{U}}{\rm{diag}}\!\left( {{{{\psi _1}{{\left( {{\lambda _1} \!+\! 
		\sigma 
								_{\bf{h}}^2} \right)}^2}} \over {{{\left( 
								{{\psi _1}\left( {{\lambda _1} 
									\!+\! \sigma _{\bf{h}}^2} \right) \!+\! 
									{\sigma ^2}} \right)}^2}}}, \cdots 
			,{{{\psi _M}{{\left( {{\lambda_M} \!+\! \sigma _{\bf{h}}^2} 
			\right)}^2}} 
				\over {{{\left( {{\psi _M}\left( {{\lambda _M} \!+\! \sigma 
				_{\bf{h}}^2} 
								\right) \!+\! {\sigma ^2}} \right)}^2}}}} 
								\right)\!{{\bf{U}}^H} \\ &
		= {\bf{U}}{\bm \Lambda}_{\bm \Xi}{{\bf{U}}^H},
	\end{align}
	where ${\bm \Lambda}_{\bm \Omega}$ and ${\bm \Lambda}_{\bm \Xi}$ are both 
	$M$-dimensional diagonal matrices. By substituting the above ${\bm 
		\Omega}{{{\bf{\hat \Sigma }}}_{\bf{h}}}$, ${{{\bf{\hat \Sigma 
		}}}_{\bf{h}}}{\bf{\Xi }}{{{\bf{\hat \Sigma }}}_{\bf{h}}}$, and ${\bf 
		\Sigma_h} 
	= \mb{U}{\bf \Lambda}\mb{U}^H$ into $\hat \delta$ in 
	(\ref{eqn:useful_hat_delta}), we obtain
	\begin{align}\label{eqn:hat_delta_appendix}
		\notag
		&\hat \delta  =  \mathsf{Tr}\left( {\left( {{{\bf{\Lambda 
		}}_{\bf{\Omega 
				}}} - {{\bf{I}}_M}} \right){\bf{\Lambda }}\left( {{{\bf{\Lambda 
					}}_{\bf{\Omega }}} - {{\bf{I}}_M}} \right)} \right) + 
					{\sigma 
			^2}\mathsf{Tr}\left( {{{\bf{\Lambda }}_{\bf{\Xi }}} } \right) \\
		= & {\sigma ^2}\sum\limits_{m = 1}^M {{{{\lambda _m}{\sigma ^2} + {\psi 
						_m}{{\left( {{\lambda _m} + \sigma _{\bf{h}}^2} 
							\right)}^2}} \over {{{\left( 
							{{\psi _m}\left( {{\lambda _m} + \sigma 
							_{\bf{h}}^2} 
								\right) + {\sigma ^2}} 
							\right)}^2}}}} \notag \\
		= & {\sigma ^2}\sum\limits_{k = 1}^K {\frac{{{\lambda _k}{\sigma ^2} + 
					{\psi 
						_k}{{\left( {{\lambda _k} + \sigma _{\bf{h}}^2} 
							\right)}^2}}}{{{{\left( {{\psi 
									_k}\left( {{\lambda _k} + \sigma 
									_{\bf{h}}^2} 
								\right) + {\sigma ^2}} 
							\right)}^2}}}}  + \sum\limits_{m = K + 1}^M 
		{\frac{{{\psi _m}\sigma^2\sigma 
					_{\bf{h}}^4}}{{{{\left( {{\psi _m}\sigma _{\bf{h}}^2 + 
					{\sigma 
									^2}} 
							\right)}^2}}}},
	\end{align}
	where the second equality holds since $\lambda_m=0$ for all $m>K$. 
	According to 
	(\ref{eqn:WWH}), for the water-filling estimator, ice-filling estimator, 
	and 
	random estimator, we can replace the $\psi_m$ in 
	(\ref{eqn:hat_delta_appendix}) 
	with ${\hat p}_m$, ${\hat n}_m$, and $\frac{Q}{M}$ to obtain ${{\hat \delta 
		}_{{\rm{wf}}}}$, ${{\hat \delta }_{{\rm{if}}}}$, and ${{\hat \delta 
		}_{{\rm{rnd}}}}$, respectively. According to 
	(\ref{eqn:hat_delta_appendix}), it is obvious that $\hat \delta \mathop  = 
	\limits^{Q \to  + \infty } {\cal O}\left( {{\sigma ^2}{M^2}{Q^{ - 1}}} 
	\right)$ holds for the three estimators. This completes the proof.
\end{appendices}

\footnotesize
\bibliographystyle{IEEEtran}

\bibliography{IEEEabrv,reference}	


\end{document}